\pdfoutput=1

\documentclass[aps,showpacs,preprintnumbers,amsmath,amssymb, 
eqsecnum,
twocolumn, tightenlines
]{revtex4}

\usepackage{graphicx}

\newcommand{\be}{\begin{eqnarray}}
\newcommand{\ee}{\end{eqnarray}}

 \newcommand{\gsim}{\mathrel{\hbox{\rlap{\lower.55ex \hbox {$\sim$}}
                   \kern-.3em \raise.4ex \hbox{$>$}}}}
\newcommand{\lsim}{\mathrel{\hbox{\rlap{\lower.55ex \hbox {$\sim$}}
                   \kern-.3em \raise.4ex \hbox{$<$}}}}

\def\roughly#1{\mathrel{\raise.3ex\hbox{$#1$\kern-.75em%
\lower1ex\hbox{$\sim$}}}}
\def\lsim{\roughly<}
\def\gsim{\roughly>}
\def\fm{{\mbox{fm}}}

\begin{document}


\title{ The
 Fate of the Initial State Perturbations 
in Heavy Ion Collisions }
\author {Edward Shuryak}
\address {Department of Physics and Astronomy, State University of New York,
Stony Brook, NY 11794}
\date{\today}

\begin{abstract}
Heavy ion collisions at RHIC are  well described
by the (nearly ideal) hydrodynamics. In the present paper we study
propagation of perturbations induced by moving charges (jets) on top of the expanding fireball, 
using hydrodynamics and (dual) magnetohydrodynamics.
Two experimentally observed structures, called a ``cone'' and a ``hard ridge'',  have been discovered
in dihadron correlation function with large-$p_t$ trigger,
while ``soft ridge'' is a similar structure seen without hard trigger. 
All three can be viewed as traces left by a moving charge in matter, 
on top of overall expansion. A puzzle 
is why those perturbations are apparently rather well preserved
at the time of the fireball freezeout.
We study two possible solutions to it: (i) a ``wave-splitting'' acoustic   
option and (ii)  a ``metastable
electric flux tubes''. In the first case we show that rapidly  variable speed of sound  under certain conditions
 leads to secondary sound waves, which are
at freezeout time closer to the original location
and have larger intensities than the first wave. 
 In the latter case we rely on (dual) magnetohydrodynamics, which also predicts two cones or cylinders of the waves.
We also briefly discuss metastable electric flux tubes in the near-$T_c$ phase and their relation to clustering data. 
\end{abstract}
\maketitle

\section{Introduction}


  The  issues to be discussed are somewhat similar in nature to what happened in cosmology in the last decades. While
  the average Hubble-like expansion of the Universe has been dramatically confirmed by the discovery of background radiation
 more than 40 years ago,  more recent observations of small-amplitude temperature
fluctuations have transformed cosmology into a much more quantitative science.

Similarly, experimental data obtained in heavy ion collisions at RHIC were shown to be in very good agreement with 
 hydrodynamical description of the ``Little Bang". Especially good results are obtained in 
 hydrodynamics supplemented by the hadronic cascades 
\cite{hydro,Hirano:2004ta,Nonaka:2006yn}. Dissipative effects  from viscosity provide only small corrections, at the few percent level,
see more in  \cite{Romatschke:2007mq,Dusling:2007gi,Heinz:2008qm}. 
Except for rather short time of initial accelration, the hydro solution can actually be rather well approximated by
 Hubble flow 
$ v(t,r)=H r $ 
with $H\approx 0.08 \, fm^{-1}$ being approximately
space and time-independent . If so, the expansion can be approximated by a quite simple form
\be  r(t)=r(0) exp(H t) \ee
which we will use below.

In the last few years RHIC experiments have focused more on two and three-particle correlations, which revealed rather rich phenomenology
of correlations. These correlations appear due to  certain fluctuations, propagating on top of the overall Hubble-like expantion.
Quite puzzling dynamics of such perturbations  is the subject of this paper.
%
%
%
We will turn to experimental observations in the next section: but before we do so, let me formulate the main
  dilemma of this work:  
%
%
 either\\ (A) these perturbations   are 
$hydrodynamical$  in nature, although propagating a bit differently
from what can be naively expected on the basis of a geometric optics, \\  or (B) 
they are {\em  not  hydrodynamical} but include certain extra fields/structures, affecting
 their expansion.

In this work we will examine whether both of those solutions
are viable. The option (A) -- to be referred to as 
the {\em ``acoustic solution"}
-- will reveal creation of the secondary  waves, induced by time-dependent
speed of sound. (In fact this effect was already noticed
  in Ref.  
 \cite{CasalderreySolana:2005rf} in connection with conical flow.)
 As we will show,  such secondary waves are brighter
and smaller in size, as sketched in Fig.\ref{fig_sketch}(c). However, as we will find,
it is not clear whether solution A will be viable quantitatively, 
as it require rather sharp drop in a speed of sound.

The second option B also leads to double cones, now as two components of Alfven waves in a (dually)magnetized medium.
Furthermore, some of them have small or even zero expansion velocity, and indication to existence of stabilized {\em  
electric flux tubes}  
in near-Tc temperature interval. Metastable microscopic flux tubes in the near-$T_c$ region  had also been considered in a different context
before, by Liao and myself  \cite{Liao:2007mj,Liao:2008vj} in connection with lattice data on lattice potentials and charmonium survival.
Yet again, 
although such tubes  have good reasons to exist, the final conclusion
on whether they are robust enough to explain the observed
``cone'' and two ``ridges'' would require a lot of further
experimental and theoretical work.

  
Early stages of heavy ion collisions are believed to be described by the
so called ``glasma" ,  a set of random color fields 
created by color charges of partons of  the two colliding nuclei
at the moment of the collision 
 \cite{McLerran:1993ni}. 
  For large nuclei those charges and fields can become large enough to be treated classically.
However as two discs with charges move away from each other, those classical field are getting smaller and (in a still
poorly understood process) rather quickly create the quark-gluon plasma, in which the occupation
numbers becoming $O(1)$.
 
 Perturbative theory of  asymptotically  hot QGP predicts  
 perturbative electric screening mass $M_E\sim gT$ from the one-loop perturbative
polarization tensor 
 \cite{Shuryak:1977ut}. However perturbative approach provides no screening of the static magnetic fields, $M_M=0$, like in the QED plasma.
%
%
  The crucial difference between the QED and QCD plasmas
 lies in the existence of magnetically charged quasiparticles -- monopoles and dyons --
leading to nonzero magnetic screening mass $M_M\sim g^2T$ first suggested by Polyakov 30 years ago \cite{Polyakov:1978vu}
and by now well confirmed by subsequent lattice studies.
Thus hot QGP, unlike the electromagnetic plasmas,  screen $both$ the electric and the magnetic fields at some microscopic scales, although at a bit 
different ones.   

  Furthermore,  recent theory developments known as ``magnetic scenario" \cite{Liao_ES_mono,Chernodub:2006gu} for the near-$T_c$
  region show that
  the situation with electric and magnetic screening masses get in fact inverted in this region. Multiple lattice studies, e.g.
ref \cite{Nakamura:2003pu} 
 have shown that at  $T<1.4T_c$ the relation between electric and magnetic screening masses get inverted, $M_M>M_E$.
 As $T\rightarrow T_c$  the
       electric screening mass strongly decreases, partly because of heavy quark and gluon quasiparticles and partly because
       of their suppression by small  Polyakov loop expectation value $<L>$, going to zero below $T_c$. As temperature decreasing toward $T_c$
 the magnetic screening mass is  instead increasing monotonously, to a rather large value
 \be M_M(T\rightarrow T_c)\approx 3T_c,  M_E(T\rightarrow T_c)\approx 0 \ee      
Such behavior of  screening as well as other theoretical considerations had lead to the ``magnetic scenario" \cite{Liao_ES_mono,Chernodub:2006gu},
which basically views the near-$T_c$ region ($0.8T_c$ up to 1.4$T_c$) as magnetic-dominated plasma, dominated by (gluomagnetic) monopoles/dyons. 
(It was formally known as the ``M-phase", from ``mixed" of the 1st order transition, can 
 now also be called M-phase from ``magnetic". ) Furthermore, even ``electric-dominated"
QGP, at $T>1.4T_c$, has viscosity and diffusion strongly influenced by electric-magnetic scattering
 \cite{Liao_ES_mono,Ratti:2008jz}.
These effects  provide a natural explanation for unusually small
viscosity observed at RHIC and also predicting its value at the LHC. 

\begin{figure}[t]
\begin{center}
\includegraphics[width=7cm]{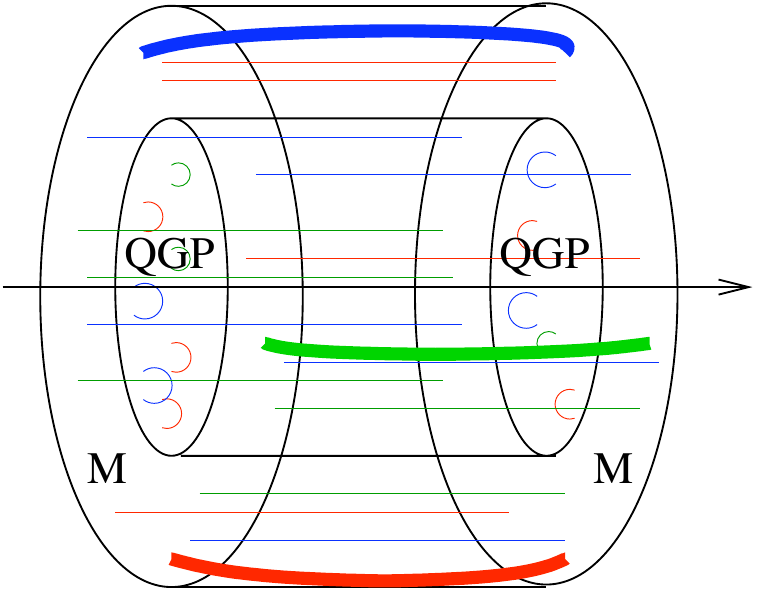}
\includegraphics[width=7cm]{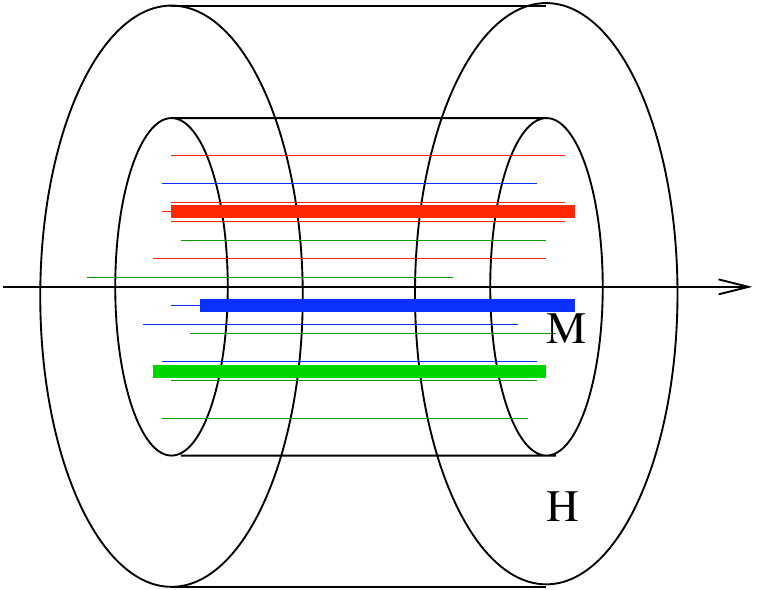}
\end{center}
\caption{\label{fig_dual_mag} 
A snapshot of unscreened electric (dual-magnetic)
field in the M (near-$T_c$) region of the fireball. Fig.(a) and (b) correspond to
full RHIC energy and the reduced energy (analogous to SPS).}
\end{figure}

The most important consequence of that for the present paper is that in the M-phase of the collision 
 the magnetic field is well screened while the  $electric$ one remains (nearly) $unscreened$.
This leads to the central new idea of this paper: that QGP produced in central part of RHIC collisions should have a ``dual corona" in which $electric$
(rather than magnetic) fields coexist with the plasma, affecting both the overall expansion and the propagation of perturbations.  
 We will suggest to use ``dual magnetohydrodynamics" (DMHD) for the description of diffuse electric fields in the M-phase, in particular
 study their effect
on the velocity of propagation of small perturbations. We will further argue below that like solar corona, that of QGP  should have metastable  flux tubes, although 
 microscopically thin ones which cannot be directly described by DMHD approximation. 


The word ``corona" used here comes from the physics of the Sun. Let me briefly remind the reader that 
it was started by Galileo Galilei, who in 1612 spent some time observing the motion of the black spots on the Sun and correctly concluded from  motion
 of the spots that they must resign on a surface of a rotating sphere: he thus argued the spots were not 
 shadows of some planets passing in front, as it was thought of before. In due time relation between the spots and solar magnetism
 was understood: modern telescopes allows one to see the
fine structure of solar spots, resolving  individual magnetic flux tubes. 
Better understanding of solar magnetism came with the advance of plasma physics in 1940's and development of MHD, which explained both the influence of
diffuse magnetic field on plasma and formation and 
 mechanical stability of the flux tubes. 
The MHD flux tubes are supported by the electron current, while the positive charges -- the ions -- are heavy and dont move. Since it is not a supercurrent,
there is inevitable friction and thus metastability of the flux tube solutions. 


As at RHIC the central part of the produced fireball
reaches relatively high temperature $T\sim 2T_c$, we expect both $E,B$ fields to be effectively screened there,
see the central cylindrical part of Fig.\ref{fig_dual_mag}  marked QGP.
But in the near-$T_c$ 
region  vanishing electric component leads to vanishing electric screening mass. 
This means that plasma in the outer  
cylindrical part of Fig.\ref{fig_dual_mag}  marked M (mixed or magnetic) is nearly pure magnetic.
It is very important to emphasize that although this region on the phase diagram
is represented by a very narrow strip  $|T-T_c|\ll T_c$, it corresponds to more than order of variation
 of the energy or entropy density, and the corresponding space-time volume
 in the expansion of the fireball is by no means small. I
A snapshot of the geometry of the M region at  some early time  is shown in Fig.\ref{fig_dual_mag}: here are unscreened electric fields (thin lines) and
magnetic flux tubes (think lines).
The lower plot show similar snapshot at collision energy much smaller than at RHIC, planned to be investigated in a specialized run.

\section{New  structures observed in two and three particle correlations}
\subsection{The cone and the ridges}
\label{sec_exper}
 Three different correlation phenomena have been discovered in
 heavy ion collisions at RHIC:\\
(i) the so called ``cone'' \cite{STAR_cone,PHENIX_cone}, is a two-peaked structure
seen in azimuthal distribution of hadrons on the ``away-side'' from a trigger hadron
(the region off quenched companion jet);\\
(ii) the so called ``hard  ridge'' seen on the ``same-side'' 
 in the triggered events
\cite{Putschke}; \\
(iii) and the ``soft ridge'' observed in 2-particle correlations 
without any hard trigger \cite{Adams:2005aw} in the minijet region,
with transverse momenta $p_t\sim 1-2 \,GeV$.

\begin{figure}[h]
\begin{center}
\includegraphics*[width=8.cm]{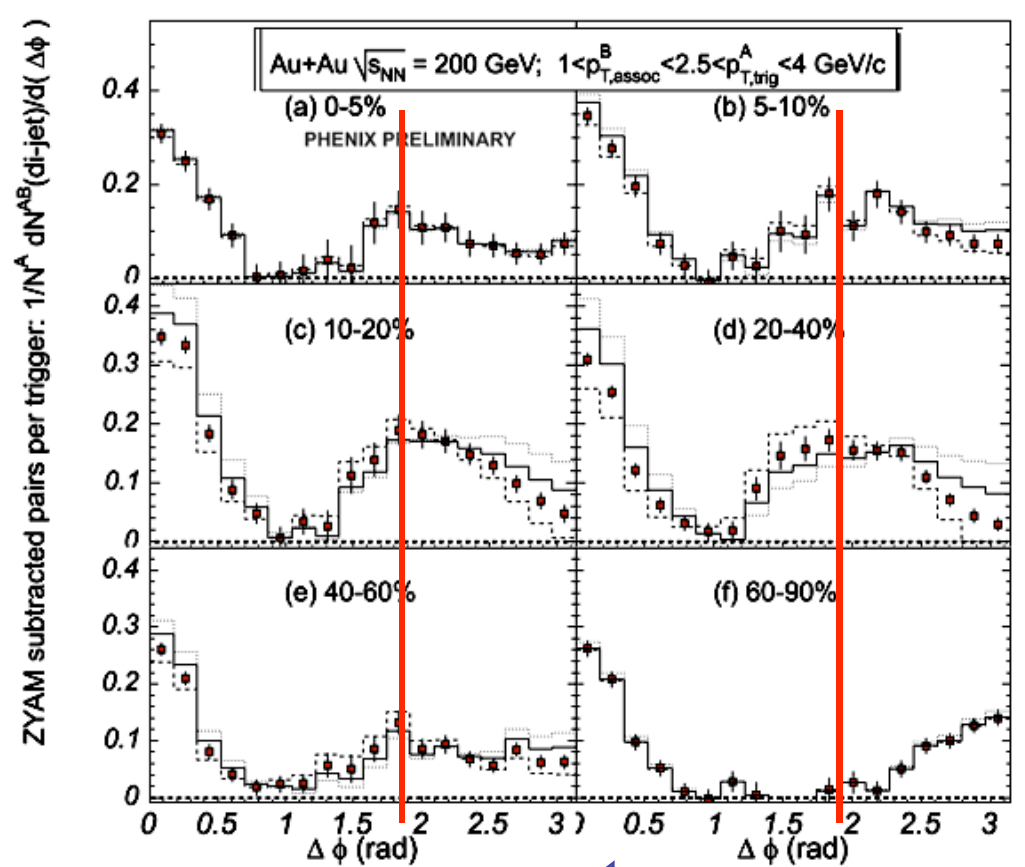}
\caption{A set of two-particle correlators from PHENIX collaboration,
as a function of azimuthal angle difference $\Delta \phi$.
Six pictures are for different centrality classes, indicated by
percentage of the total cross section.
While the most peripheral collisions (the right-lower corner) shows two peaks, one near zero and one near
$\phi=\pi$, as in pp collisions, others show a minimum for that ``away-side" angle and a peak
shifted by a large angle from it (vertical solid lines).
}
\label{fig_cone}
\end{center}
\end{figure}

(i) The ``cone'' has been discovered in the 2-particle azimuthal
correlations like the one shown in Fig.\ref{fig_cone}.
 One can see from this figure
the disappearance of the ``away-side'' peak
 at $\Delta\phi=\pi$ and appearance of new peaks at completely
 different angle, as one moves from peripheral to central collisions.
After discovery of those effects there was extensive studies of the 3-particle
correlations as well. This is  a rather complicated subject to go into here, let me just say that
they has confirmed the observed structure is indeed cone-like,  and not e.g. a reflected jet.

(ii) the hard ridge is also seen in 2-particle correlators,
but plotted on the two-dimensional $\Delta \phi-\Delta \eta$ plane,
the differences between the azimuthal angles and pseudorapidities
 of the two
particles.
The
 jet remnants make a 
peak  near   $\Delta \phi=0,\Delta \eta=0$, which was found to
sit on top of
the ``ridge'', with  comparable width in $\Delta \phi$ but
 very wide width $\Delta \eta$.
 For plots and various features one can consult
the original talk by  Putschke \cite{Putschke}. Later it
was shown by PHOBOS collaboration \cite{phobos} that the rapidity range of the ridge
extends at least up to
$|\eta|\approx 4$.

(iii)  the ``soft ridge'' is found by STAR collaboration \cite{Adams:2005aw,star_ridge}
  without a trigger, in the 2-particle correlations.

For many experimental details and phenomenological
considerations related to these phenomena the reader may 
consult e.g.
 the talks at recent specialized workshop
 \cite{cathie_workshop}.

\begin{figure}[t]
\begin{center}
\includegraphics*[height=10.cm,width=4.cm]{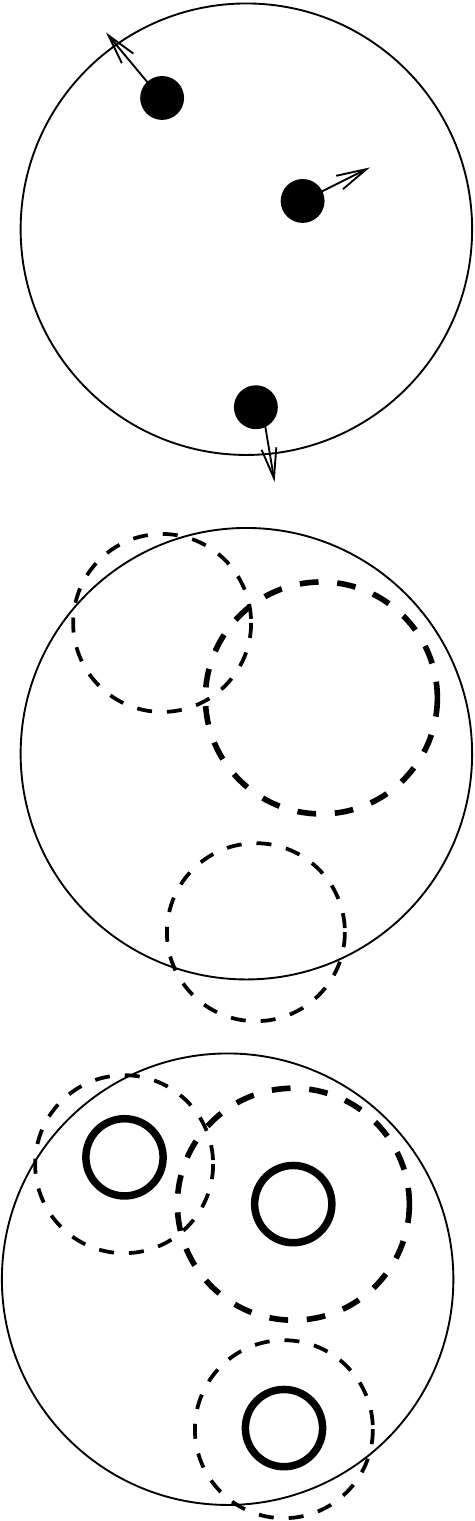}
\caption{A sketch of the transverse plane of the colliding system:
the ``spots'' of extra density (a) are shown as black disks, to be moved
by collective radial flow (arrows). Naive sound expansion (b)
 would produce large-size and small amplitude wave: yet
the correct solution includes also brighter $secondary$ wave (c)
of smaller radius.
}
\label{fig_sketch}
\end{center}
\end{figure}

 We will  return to these observations  below,
 turning now to
 their suggested explanations:\\
(i) Stoecker et al,
 as well as Casalderrey, Teaney and myself \cite{conical} have proposed 
that the energy deposited by a quenched jet goes into two
hydrodynamical excitation modes, the sound and the so called diffusion or wake modes.   
The sound from the propagating jet  should thus  create the famous Mach cone,
in qualitative agreement  with the conical structure observed. \\ 
(ii)  One early model for ``hard ridge'' has been 
introduced in my paper \cite{Shuryak:2007fu}.  It relates it with the
forward-backward jets accompanying any hard scattering,   providing
extra particles (``hot spot") widely distributed in rapidity.
This idea is then combined with the one  suggested previously by
Voloshin \cite{Voloshin:2003ud}, namely that extra particles
deposited in the fireball would be moved transversely by  the radial
hydrodynamical flow,  should produce a peak at certain
      azimuthal angle corresponding to the position of the hot spot,  see Fig.\ref{fig_sketch}(a).
While particles of the ridge are separated by large rapidity gaps and cannot
communicate during the expansion process,
their azimuthal emission angles remain correlated with each other because they originate from
 the same  ``hot spot'' in the transverse plane.\\
(iii) Similarly, transverse hydro boost of ``hot spots" was used for the
  explanation of the ``soft ridge''
  by McLerran and collaborators 
 \cite{Dumitru:2008wn,Gavin:2008ev}. They have pointed out that
the initial state color fluctuations in the
colliding nuclei would create  longitudinal ``color flux tubes'' 
 $without$ any hard collisions. 
 As these tubes are being stretched between two fragmentation regions
of the colliding nuclei,  they also lead to long-range rapidity
 correlations. 

\subsection{Naive hydrodynamics and the remaining puzzles}

So, at a very qualitative level the origin of all three phenomena seem to be explained:
yet at more qualitative level a lot of puzzles appear. As an example, consider
the simplest of them, the ``soft ridge". As discussed in  \cite{Dumitru:2008wn,Gavin:2008ev},
 the initial stage   (proper time 
$\tau\sim 1/Q_s \sim 0.2\, fm/c$ where $Q_s\sim1\,GeV$ is the so called saturation scale
at RHIC) can be discussed using classical Yang-Mills
equations: thus color fluctuations naturally appear. 
However, the observed pions come from final freezeout time,
separated from the initial ``glasma" era by much longer time $\tau\sim 10\, fm$.
This is certainly so, as the explanation heavily relies on radial hydro velocity
and thus it has to wait till the hydro velocity is being created. As we will argue
below, there are many reasons why one might have expected nearly complete
disappearance of this signal during this time. 


Common to all three cases is deposition of 
 some additional energy (or entropy), on top of  the ``ambient matter". 
 The number of correlated particles in all of them 
constitute a 
small $(\sim 10^{-3})$ fraction of the total
multiplicity: thus they  can only be seen in a 
high-statistics correlation analysis.  Furthermore, simple estimates show that it would not be possible to detect
any trace of that tiny perturbation if it would be distributed over a significant fraction of the fireball:
the only possibility is that it remains well localized in transverse direction. 

 Smallness of perturbation in respect to total system size by itself does not guarantee that the
perturbations  is small $locally$, in respect to local
density of ambient matter. However it will become so if perturbation would give rise to
divergent conical (or cylindrical, or spherical)  waves, see Fig.\ref{fig_sketch}(b). 
Similar to circles from
a stone thrown into a pond, initial
 perturbation may
 become some waves, with basically nothing left
  at the original location at later time.  Even without dissipation, ideal hydrodynamics
  predicts that
the final radius of those waves is given by
 the  ``sound horizon''
\be R_{h}=\int_0^{\tau_f} d\tau c_s(\tau)  \ee
As we will detail below, 
by the  the freezeout proper time $\tau_f\sim 10-15\, fm/c$,
this distance is not small, $\sim 6 fm$ or so, since the speed of sound changes between
$c_s=1/\sqrt{3}\approx .58$ in QGP and about $.3$ at its minimum near $T_c$
 The amplitude of the wave is decreasing accordingly, and the width of $\phi$ distribution grows,
making us wandering if any trace of the perturbation can remain observable. 

And yet, we $do$ observe all three correlations, as if nothing
happened to them during rather long time of the hydro process,
$\sim 10 fm/c$. 
This is the puzzle  discussion/resolution of which
is the main objective of this paper.  The idea behind it is that there can be some reasons
providing second unusual mode of propagation, with reduction or maybe even vanishing of the speed of its spread and lead to 
an observable structures at late time, see see Fig.\ref{fig_sketch}(c).

(In the case of a cone, additional consideration is that the ``wake'' mode, behind
the
jet,  is -- in contrast -- not expanding or  weakening: and yet it is $not$ observed.
In the case of ridges, large size of waves comparable to nuclear radius will
make the radial flow directions  be rather different at
different places, widening the  peak in azimuth well beyond what is actually observed.
The puzzle is especially clear in the case
of ``ridges'', whose explanation heavily rely on substantial
hydrodynamical flow velocity, which cannot be formed promptly and
is known to be developed only by the freezeout time.)

Let me now add 
 few more important observations on the soft ridges.
The
  spectra of particles in cone and ridges,
 as well as  their  composition (not shown)
 are drastically different from
 jet remnants \cite{Putschke}.
 Particularly telling is large baryon/meson ratio which clearly
 indicate that their existence is related to the ambient matter boosted by
the hydrodynamical flow.    The boosted baryons and sharpening
of the $\phi$ peak nicely confirm  that the particles of the ridge
do come late, from the final freezeout of the system.

The fact that the cones and ridges 
  are best seen for secondaries with $p_t=1-2 GeV$ is also a confirmation of their hydro origin. The famous elliptic
flow also is maximal at such momenta, as is the baryon/meson ratio
following from the radial flow. Hydro effects in general are increasing with $p_t$ and thus are maximal at the
 upper limit of hydro description, which  is exactly
 in this $p_t$ region, as viscosity corrections tell us.

\begin{figure}[h]
\begin{center}
\includegraphics*[width=6.cm]{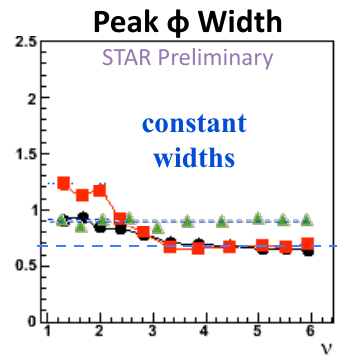}
\caption{Azimuthal  width of the ridge as a function of centrality, from the talk of L.Ray 
\protect\cite{cathie_workshop}. The red squares and black dots are for
200 and 62 GeV Au Au collisions.  Green triangles correspond to  the minijet
model without matter effects shown for comparison.
}
\label{fig_peakwidth}
\end{center}
\end{figure}

Further confirmation of hydro origin of ridges comes from the
centrality dependence of the {\em angular width} of the
ridge: the peak in azimuth $sharpens$ for
more central collisions, see Fig.\ref{fig_peakwidth}.
This happens because of two interrelated effects, 
both well documented. For central collisions
there is  (i) an $increase$ of the radial hydro velocity,
accompanied by (ii)  a substantial $decrease$ in the freezeout
temperature (which goes from  $T_c\approx 170 MeV$ in peripheral down to $T_f\approx 90 MeV$ for central
collisions.

\begin{figure}[h]
\begin{center}
\includegraphics*[width=7.cm]{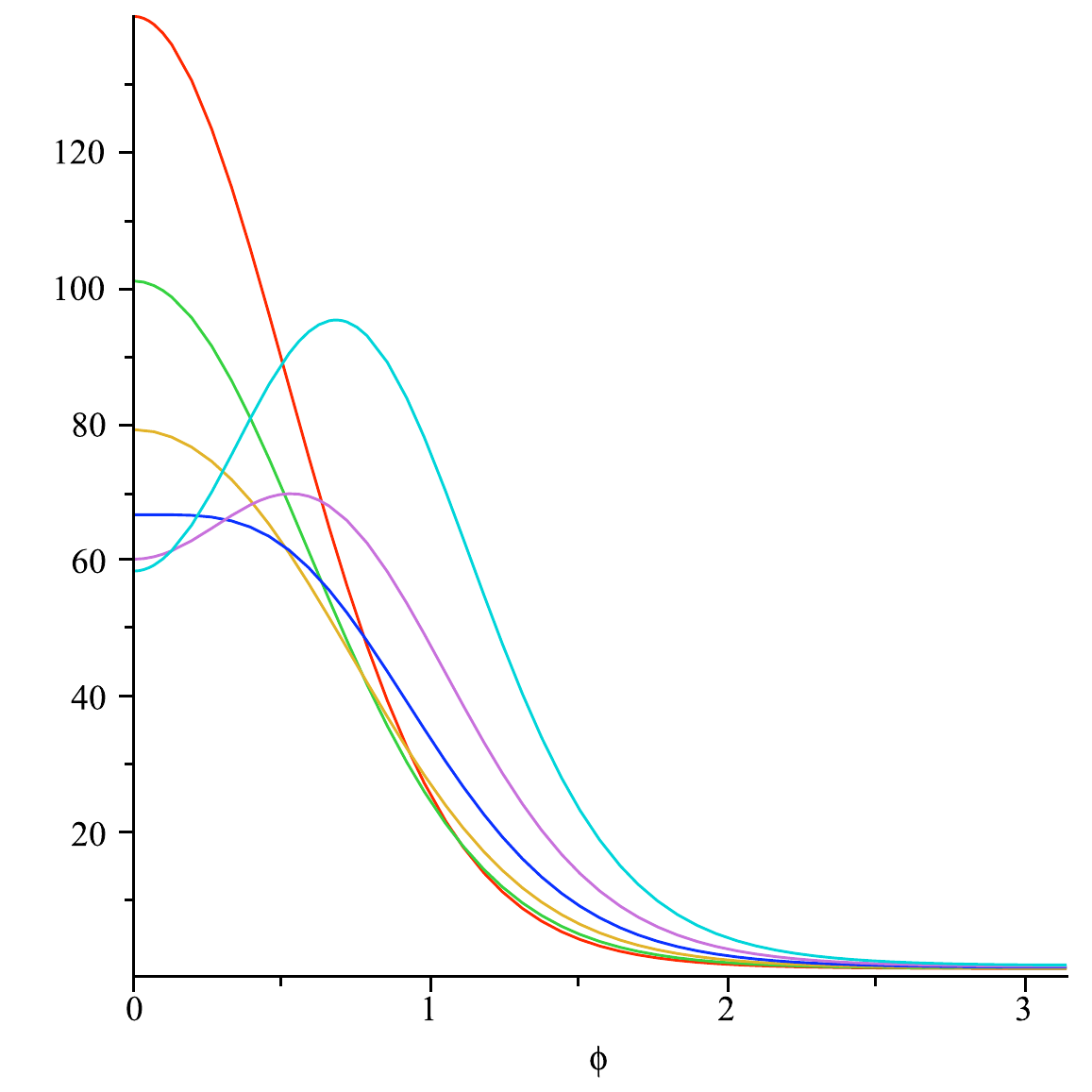}
\caption{The dependence of the visible distribution in azimuthal angle on the width of the (semi)circle 
at the time of freezeout. Six curves, from the most narrow to wider ones, correspond to the radius of the circle 1,2,3,4,5,6 fm, respectively.
The original spot position is selected to be at the edge of the nuclei.
The distribution is calculated for particle of $p_t=1\, GeV$ and fix freezeout $T_f=165\, MeV$.  
}
\label{fig_spot width}
\end{center}
\end{figure}

    Now, let us return to the puzzles.The observed width of the  azimuthal peaks
%
%
 provides strong limits on how large is the ``spot" at the freezeout
moment. In Fig.\ref{fig_spot width} we have plotted the shape of azimuthal peak produced by (semi) circles of radii 1..6 fm. 
To see those, one has to do a very simple calculation, superimposed the radial Hubble flow with the circular spot,
and calculated this angular distribution. As one can see from this figure, the width of the distribution grows -- it is     0.57,
                                0.56,
                                0.69,
                                0.76,
                                0.83,
                                0.89
for the 6 curves shown. Moreover, the distribution shapes become very different from that observed, with two maxima shifted from $\phi=0$ (corresponding
to direction of flow at two points at which the circle intersect the fireball boundary).  Comparing such distribution with observations,
e.g. their width with those shown in Fig.\ref{fig_peakwidth},one finds that
 the radius of a spot  at freezeout is restricted to be $R(\tau_{freezeout})< 3 \, fm$ or so. As we already argued in the Introduction,
this is already by about factor two smaller than the radius of the ``sound horizon" expected with the realistic speed of sound .
Therefore, naive picture of expanding hydro waves is in direct contradiction to data.

\begin{figure}[h]
\begin{center}
\includegraphics*[width=9.cm]{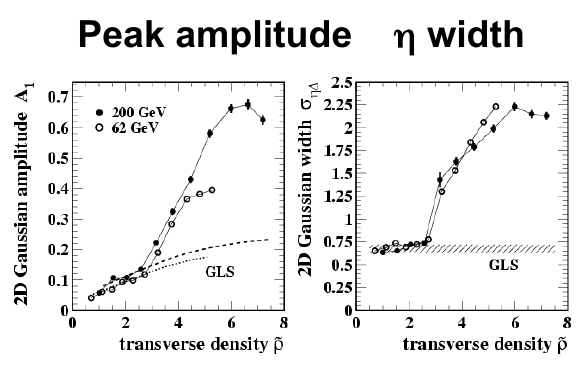}
\caption{The dependence of the amplitude and azimuthal width
of the ``soft ridge'' on the transverses multiplicity density,
 from the talk of L.Ray 
\protect\cite{cathie_workshop}. Closed and open points are for 200 and
62 GeV AuAu collisions.}
\label{fig_transition}
\end{center}
\end{figure}

Having mentioned the main puzzle, let me also point out other cases
of qualitative differences between the  overall hydro expansion and 
the (soft) ridge. The latter  has dramatic  centrality dependence
shown in Fig.\ref{fig_transition}, sharply disappearing at cenrtain centrality. (The lines and shaded area
marks GLS is some simple scaling expected from noninteracting minijet
event generator: it only describe the data at peripheral bins
 at small densities).
Furthermore,  the comparison of the
 200 and 62 GeV AuAu data shows that
the transition point seem to be at the $same$ transverse particle 
density $\rho_c=(3/2)dN_{ch}/dy/S \approx 3$, so one  may
 naively think 
 that for $\rho<\rho_c$ the matter is simply too dilute to show
hydrodynamical effects. Yet in fact both  radial and elliptic flows
  have quite smooth centrality dependence and show no rapid changes
  at the same point at all.

This difference between ridges and overall hydro flows may
be directly related to the main dilemma of this paper.
If the cones and ridges are hydrodynamical,
then why can they be so different from overall hydrodynamical 
expansion in their centrality dependence? 


  Strong temperature dependence can in principle be
related to the  issue of timing of the energy deposition.
There is a difference between
timing of the ``cone'' and ``ridges'': while the latter obviously
originate early, for ``cones''  the exact
energy deposition time/place depends on the jet quenching mechanism.
As gluon (or light quark) jets
move with a speed of light, by the time of the order of nuclear size
$\sim 6 \,fm/c$ they either leave the fireball,
 or are already completely
quenched.  Recently purely geometrical study of
angular distribution of quenching \cite{Liao:2008dk}
indicated that most of jet quenching rate should happen in the 
  near-$T_c$
 region. It is surprising, taking into account 
  much higher density of QGP at earlier time.

Another  dramatic finding, pointing to the same direction, was 
unexpected observation of similar ``cones'' by CERES and NA49
collaborations at much lower  collision energy of CERN SPS (see 
recent summary by Appelhauser in \cite{cathie_workshop}).
Since at the SPS energy there is practically no QGP
phase, it can only be there starting in the near-$T_c$ (mixed) phase.

 Obviously we would like to see what happens with the ridges
at lower collision energies. Note that both explanations
we propose in this work have problems with QGP away from
$T_c$: it is hard to stabilize the flux tube there and
also impossible to stop expansion of the sound waves with
rather large sound speed $c_s=1/\sqrt{3}$. And yet
ridges  disappear in very peripheral collisions: we would like to
know what happens
as the collision energy gets lower. (Those questions are presumably
be addressed by the expected scan down in RHIC energy, planned
in the nearest future.)

Another surprising experimental fact is quite large
 value of the cone angle, deduced from 2 and
 3-particle correlators. It seems to be
 in the range $\theta_M=1.2-1.4$ radians (not too far from $\pi/2=90^o$
or cylindrical waves!). The
Mach formula  gives the  speed of pertinent perturbation to be about 
\be <v_{wave}>=cos(\theta_M)\approx 0.2 \ee
well below the expected speed of sound (except maybe near $T_c$).
 So again, it is either (A) a coil effect, reducing expansion, or (B)
a  sound with a nontrivial production mechanism.

\section{Acoustical waves in expanding fireball with variable speed
  of sound}

\subsection{Model equation of motion}

Now we turn to discussion of the
 evolution of  small perturbations
sitting on top of overall (Hubble-like) expansion.
Equations for this case have been worked out
by Casalderrey-Solana and myself in Ref.  
 \cite{CasalderreySolana:2005rf}, and applied
to ``conical flow'' \cite{conical} from quenched jets.
 In this paper we have already found 
 that in the vicinity of the QCD phase transition there is a wave splitting phenomenon.
The framework we will use  to study the effects of the variable speed of 
sound and matter expansion 
we have looked for the simplest example possible, keeping the problem
time-dependent but
homogeneous in space. This can be 
achieved in a Big-Bang-like setting 
 in which the space is created dynamically
by gravity.
Consider a liquid in flat Freedman-Robertson-Walker metric :
\be
d\tau^2=dt^2-R(t)^2\left[dr^2+r^2(d\theta^2+sin^2 \theta d \phi)\right]
\ee
where the  parameter $R(t)$ (the instantaneous
Hubble radius of our ``universe'') is treated
as external (not to be derived from  Einstein equations
but from hydrodynamical solution.
 For isotropic expansion
only the longitudinal projection 
$u_{\mu}T^{\mu \nu}_{;\nu}$ is needed,  which is the equation
of entropy conservation leading to
\be
\label{S}
\frac{d}{dt}\left(s(t) R^3(t)\right)=0 \Longrightarrow s(t)R(t)^3=S
\ee 
for any R(t), provided expansion is adiabatic.

A
 simple substitution of a variable $c_s(t)$ 
 into the equations of motion for perturbations 
is inconsistent.
 One should instead find a correct non static solution of the
hydrodynamical equations and only then, using this solution as zeroth order,
study first order perturbations such as sound propagation.
The  linearized equations for hydrodynamical perturbations
 in this background has been derived in
 \cite{CasalderreySolana:2005rf}.
Using the normalized perturbation 
\be
\epsilon=R^4\delta T^{00} \ee
 one may eliminate
other components of the stress tensor
and get the following single equation 
\be
\label{eqn_eom}
\partial^2_{t} \epsilon - c_s^2(t) \nabla^2 \epsilon \hspace{3cm} \\ \nonumber 
+ {\epsilon \over R(t)} \partial_{t} \left[ (3c_s^2(t)-1)\partial_{t}
R(t) \right] + 
(3c_s^2(t))\frac{ \partial_{t}R(t)}{R(t)} (\partial_{t} \epsilon)=0.
\ee
 In the derivation
 we have not assumed any
particular expansion function $R(t)$ or 
particular equation of state, just general thermodynamic relations.

In  Ref.  
 \cite{CasalderreySolana:2005rf} we have used a bit different time
 variable
and  Fourier decomposition in space, reducing the problem to  
 an oscillator with the time-dependent frequency and specific
 exciting force (the third term) which is
 $negative$  for
$c^2_s <1/3$. Note that it creates amplification of dimensionless
perturbation, which is
 similar to the case of Universe expansion effect on 
small perturbations,
running away from each other.

\begin{figure}[h!]
\begin{center}
\includegraphics*[height=6.cm,width=7.cm]{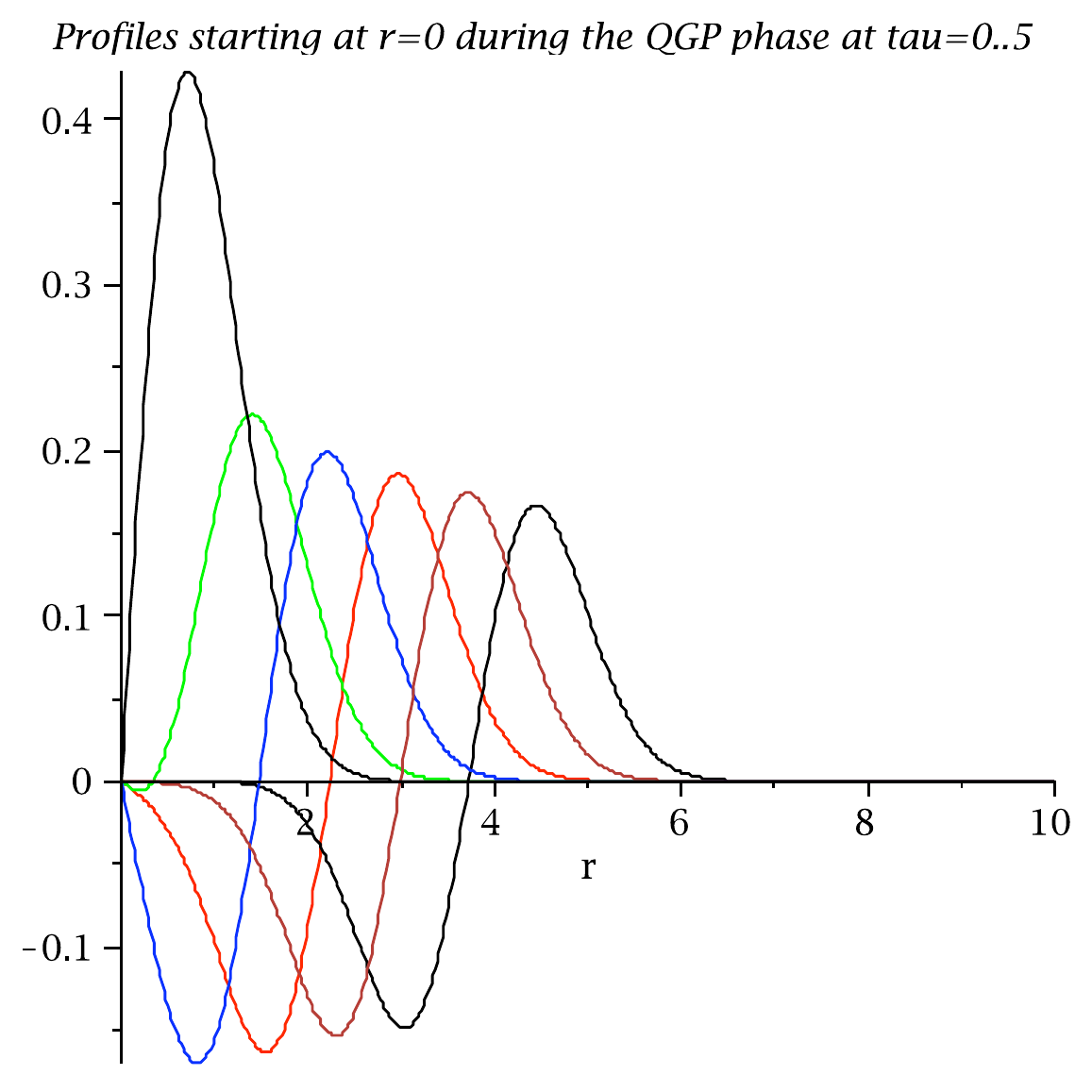}
\includegraphics*[height=6.cm,width=7.cm]{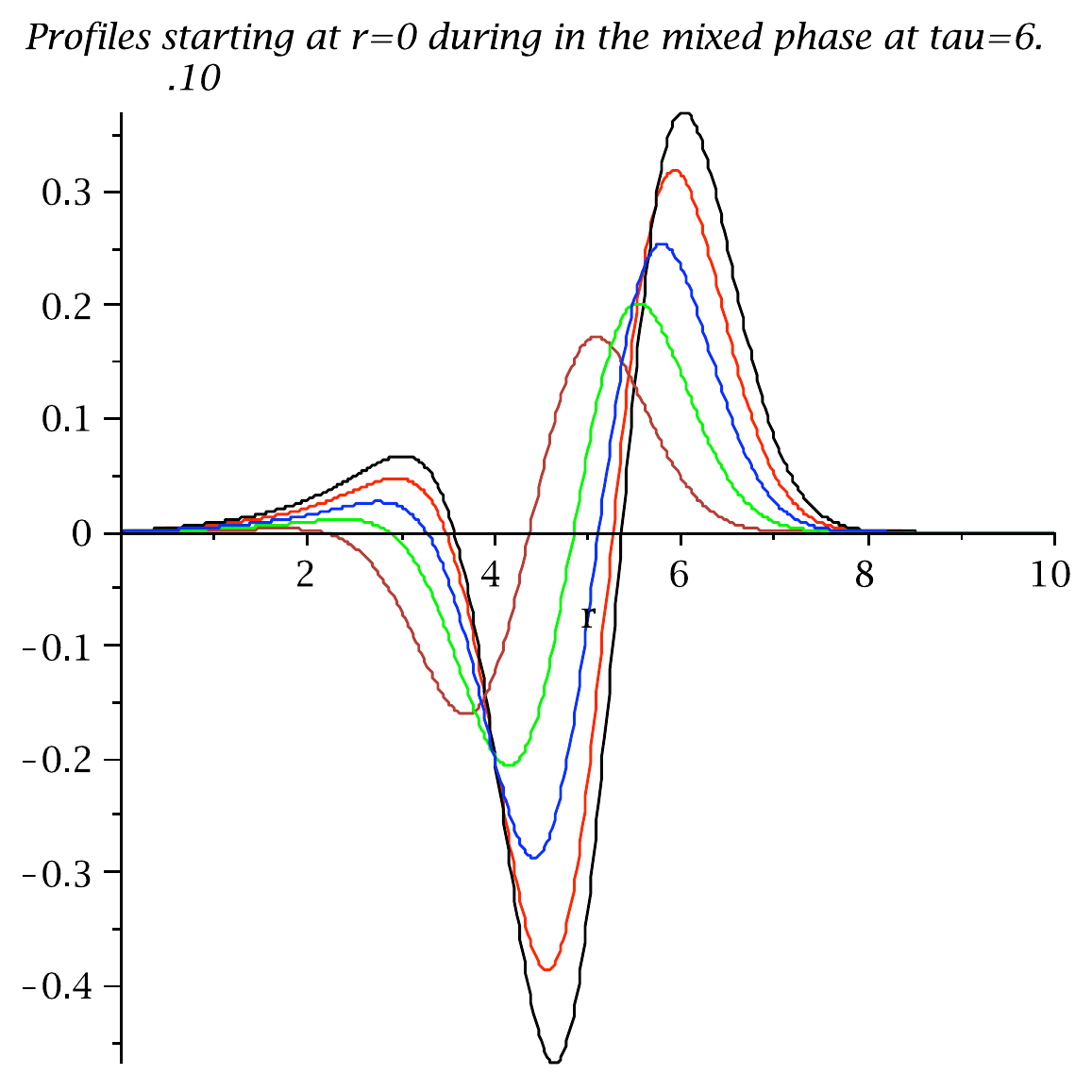}
\includegraphics*[height=6.cm,width=7.cm]{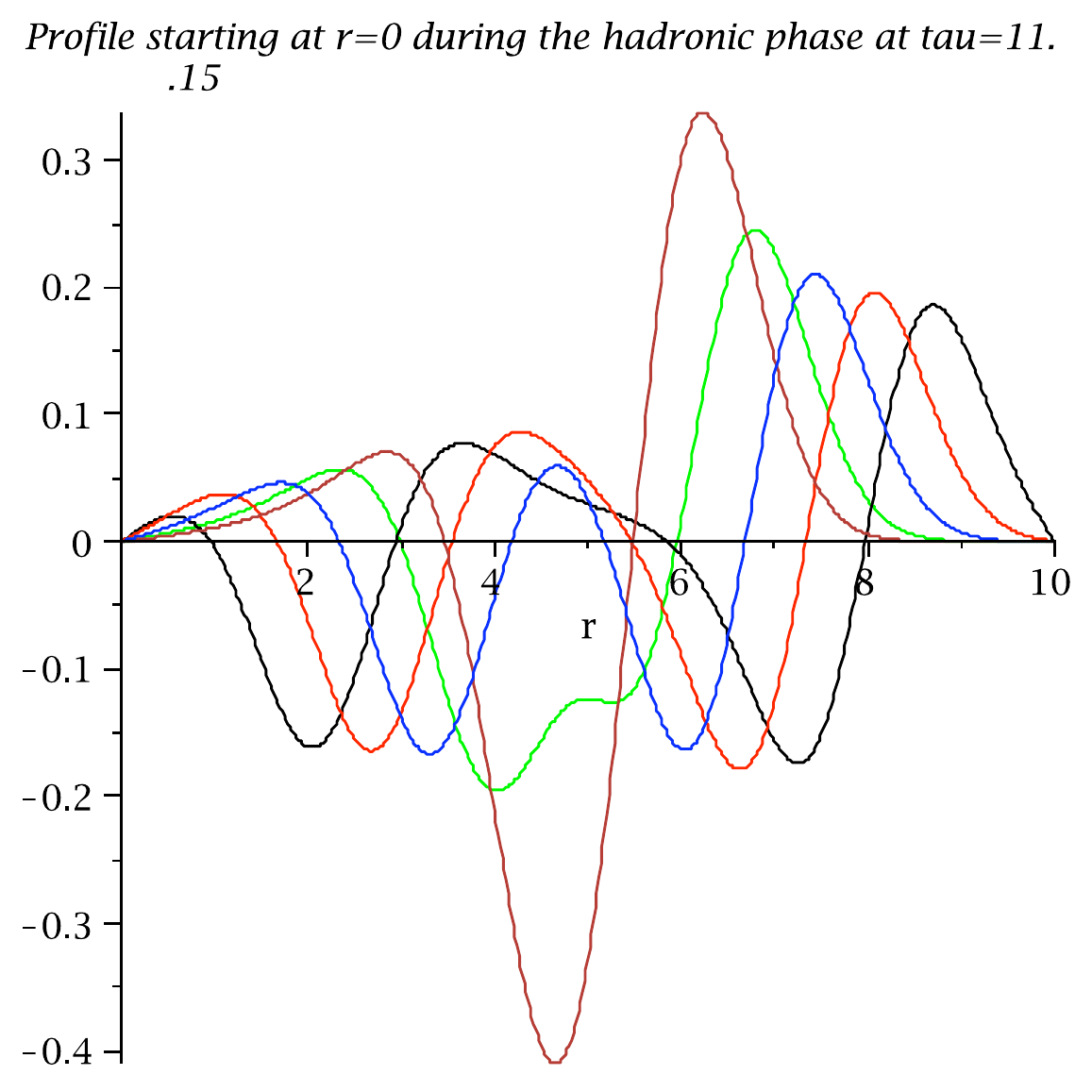}
\caption{ (Color online) The
snapshots of the evolution of perturbation
 in the QGP  (a), mixed (b) and hadronic (c) phases.
 Five curves in (a) are for subsequent wave profile at the proper time $\tau=0,1,2,3,4,5 \, fm$, in (b) for  $\tau=6,7,8,9,10 \, fm$
 and in (c) for   $\tau=11,12,13,14,15 \, fm$. The order of the lines is easily understood, as the wave moves from left to right.}
\label{fig_center}
\end{center}
\end{figure}  

\subsection{Generation of the secondary wave}

 Let us remind the setting in which the solutions were
 studied. We have already explained that we expect
the expansion to be exponential $R(t)=exp(Ht)$ and the $H$ value
was fixed from $transverse$ matter expansion. For simplicity,
we use the same expansion in 3d, although in heavy ion collisions
the longitudinal expansion is different.

\begin{figure}[t!]
\begin{center}
\includegraphics*[height=5.cm,width=5.cm]{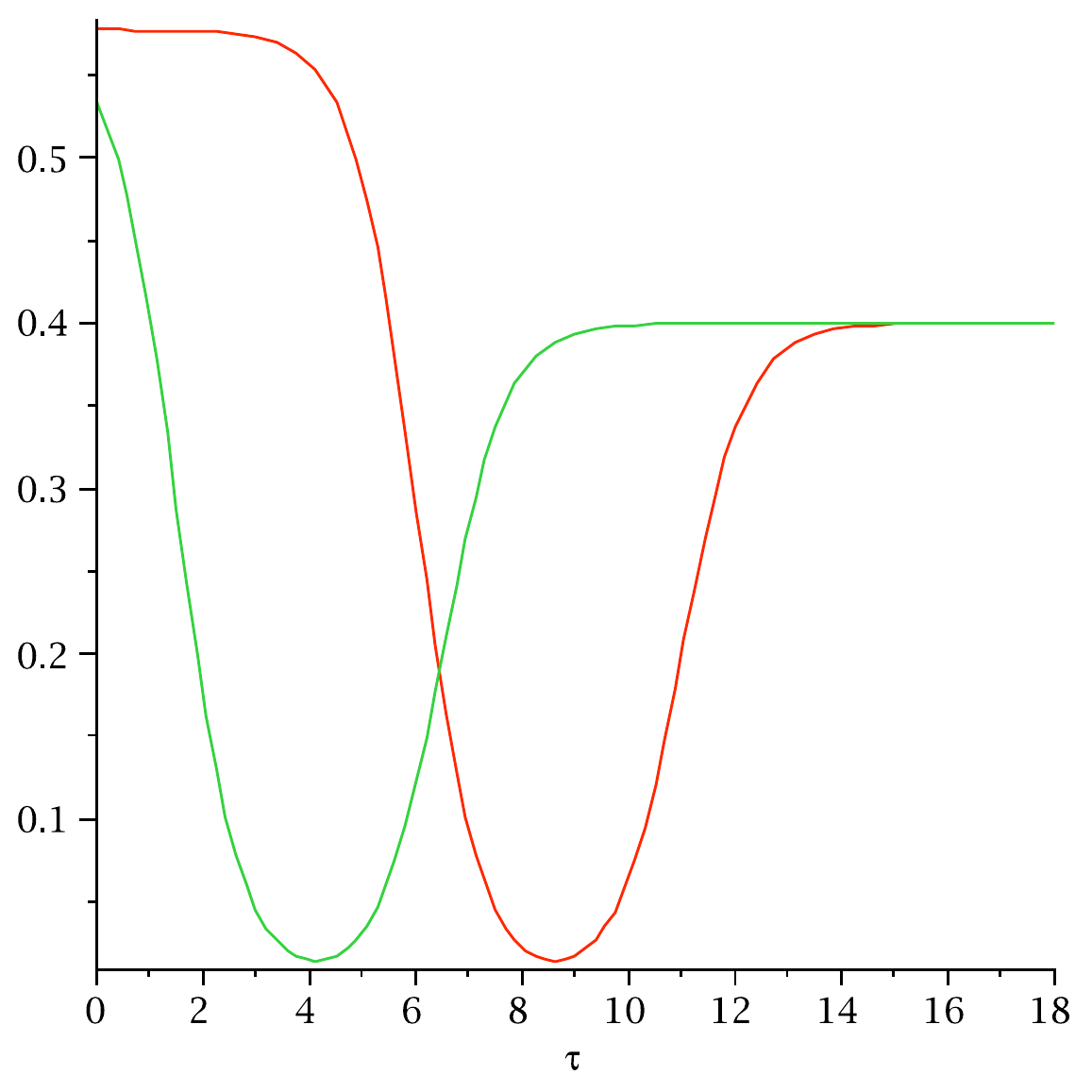}
\includegraphics*[height=5.cm,width=5.cm]{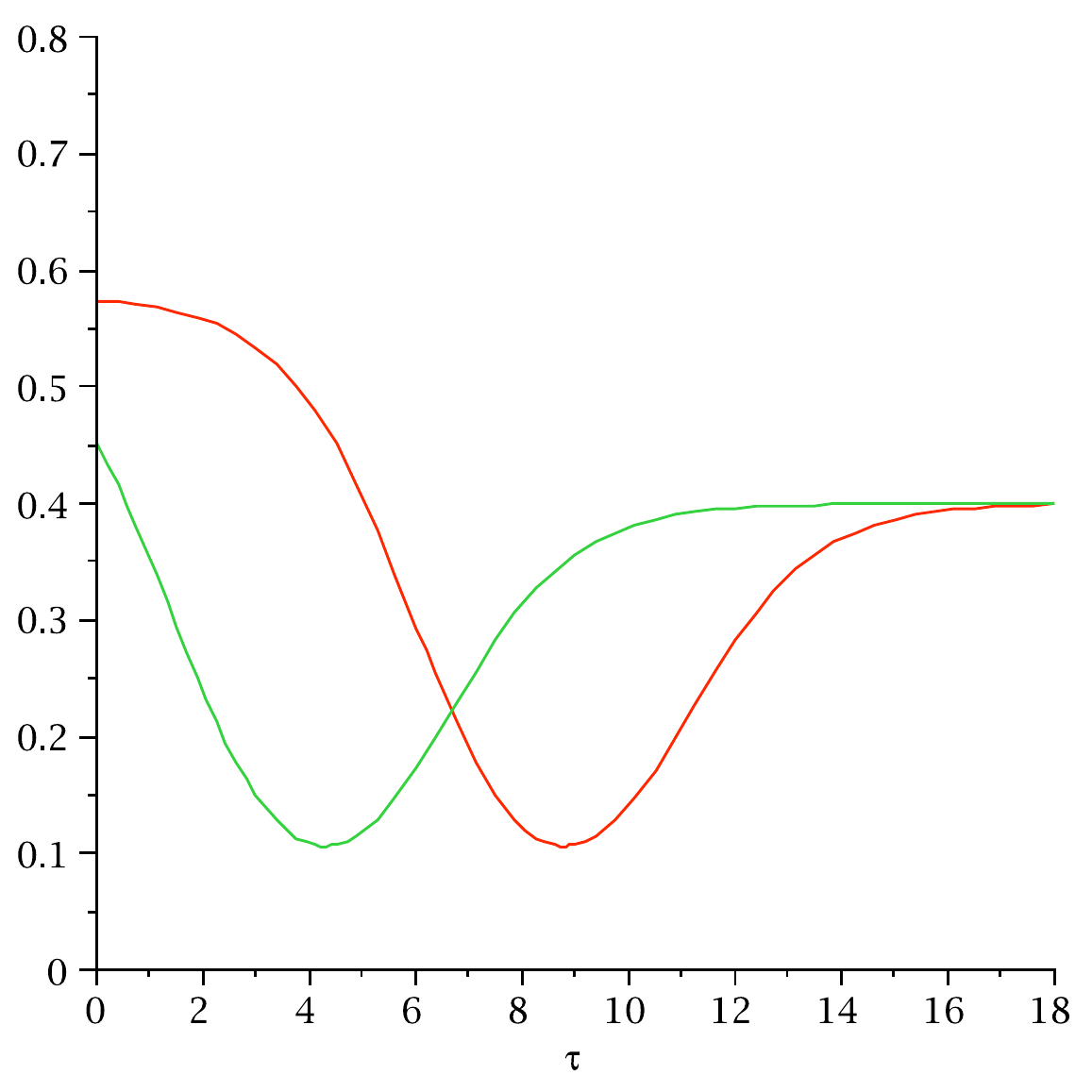}
\includegraphics*[height=5.cm,width=5.cm]{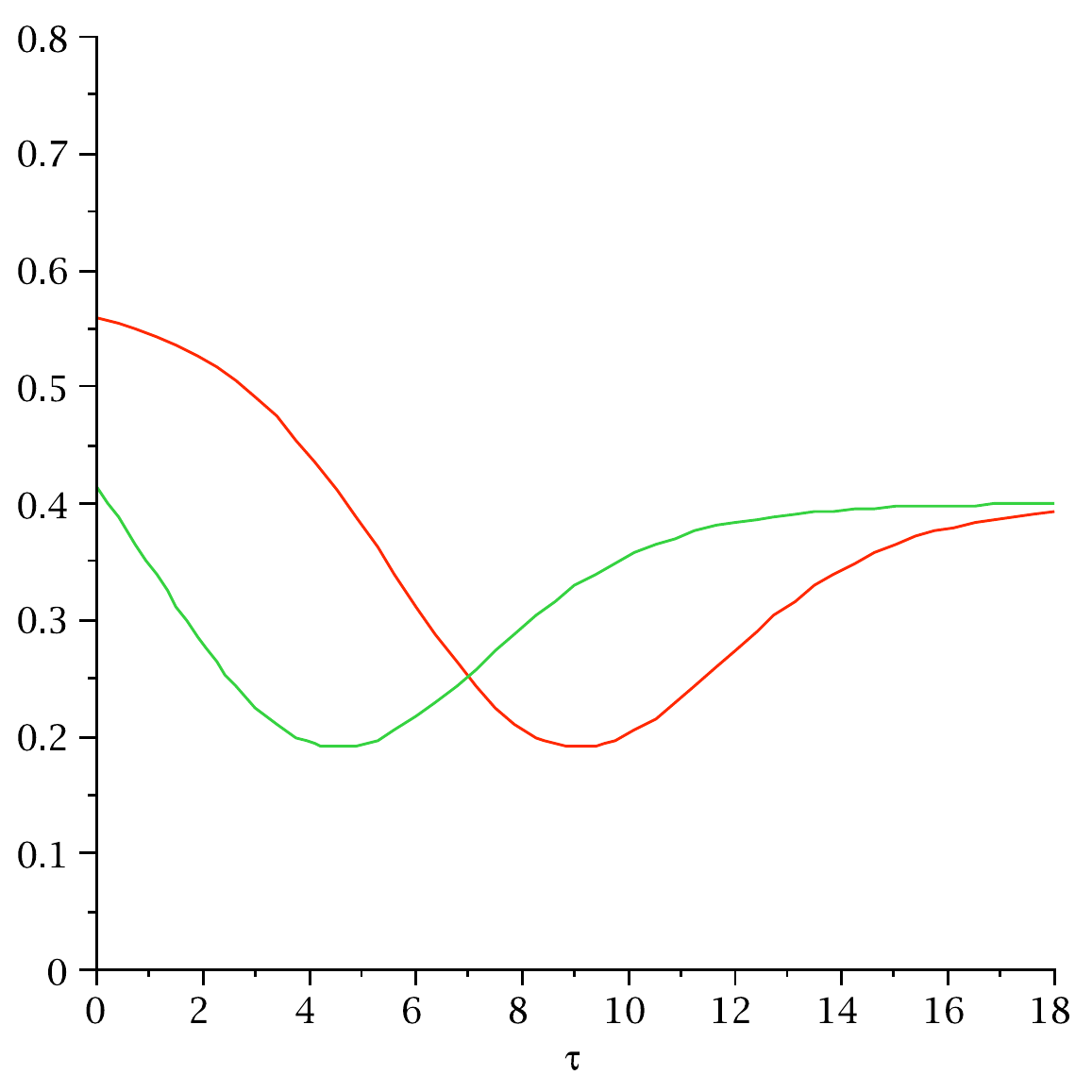}
\caption{(Color online) The sound velocity $c_s$
  as a function of proper time $\tau$
(in fm/c), for
the fireball center (r=0, red) and the rim (r=6 fm,green).
(a,b,c) are three variants of the time dependence, corresponding
to the results display in the next figure. 
}
\label{fig_cs}
\end{center}
\end{figure}

The main ingredient is the variable speed of sound $c_s(t)$.
 As we have already emphasized in the Introduction,
at early stages at
  RHIC
the matter is believed to be  in the form of
quark-gluon plasma  (QGP), and thus
with $c^2_{QGP}\approx 1/3$. This makes the third term
in our eqn (\ref{eqn_eom}):
 which is nice since in this stage the Hubble
flow is not yet a good approximation.
In the near-$T_c$ region
 the energy density is increasing much more rapidly than
 the  pressure, which makes matter ``soft'' and
 $c_s^2$ dropping to its minimum, known as the ``softest point''.
Aftre that  $c_s(t)$ is rising again, in the hadronic ``resonance
  gas''. Our model-dependent time histories at two
transverse position $r=0,6 fm$ are depicted in Fig.\ref{fig_cs},
for three different scenarios we would study.

Solutions for eqn (\ref{eqn_eom})
 corresponding to Fig.\ref{fig_cs}(a) (the sharpest
change) for initial Gaussian
perturbation
is  shown in the  Fig.\ref{fig_center}.
In each cases we show in three pictures subsequent stages of
evolution for $r*\epsilon(t,r)$ 
as profiles taken every fm/c. We start with 5 curves 
 of the fireball history, Figs.\ref{fig_center}(a)
 corresponding to the QGP era.
Since  $c^2_s \approx 1/3$ the main new (third) term 
in (\ref{eqn_eom}) is  absent,
and our solution corresponds to ``naive
evolution''  of the original  perturbation into 
 expanding primary wave. 
 No visible trace of the perturbation remains
at the initial position, and their are no secondary waves.


 In  Figs.\ref{fig_center}(b) we show what happens during
the mixed phase: the evolution of perturbations slows down
as expected. One also finds growth in amplitude, and also
the secondary wave starts to appear. Figs.\ref{fig_center}(c)
 show that  in the hadronic phase both waves
start propagating outward  -- if there is time left
to freezeout.

\begin{figure}[t!]
\begin{center}
\includegraphics*[height=5.cm,width=5.cm]{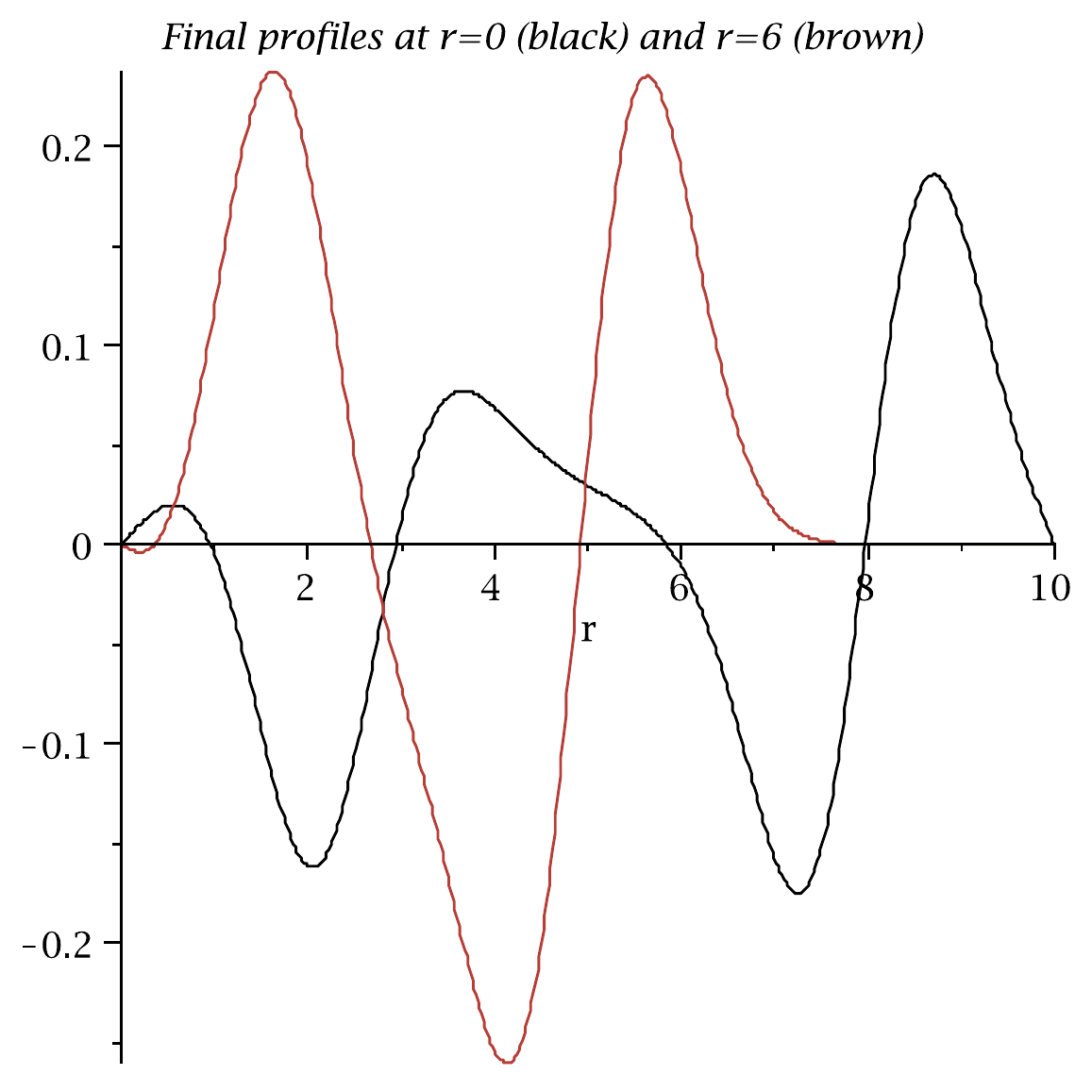}
\includegraphics*[height=5.cm,width=5.cm]{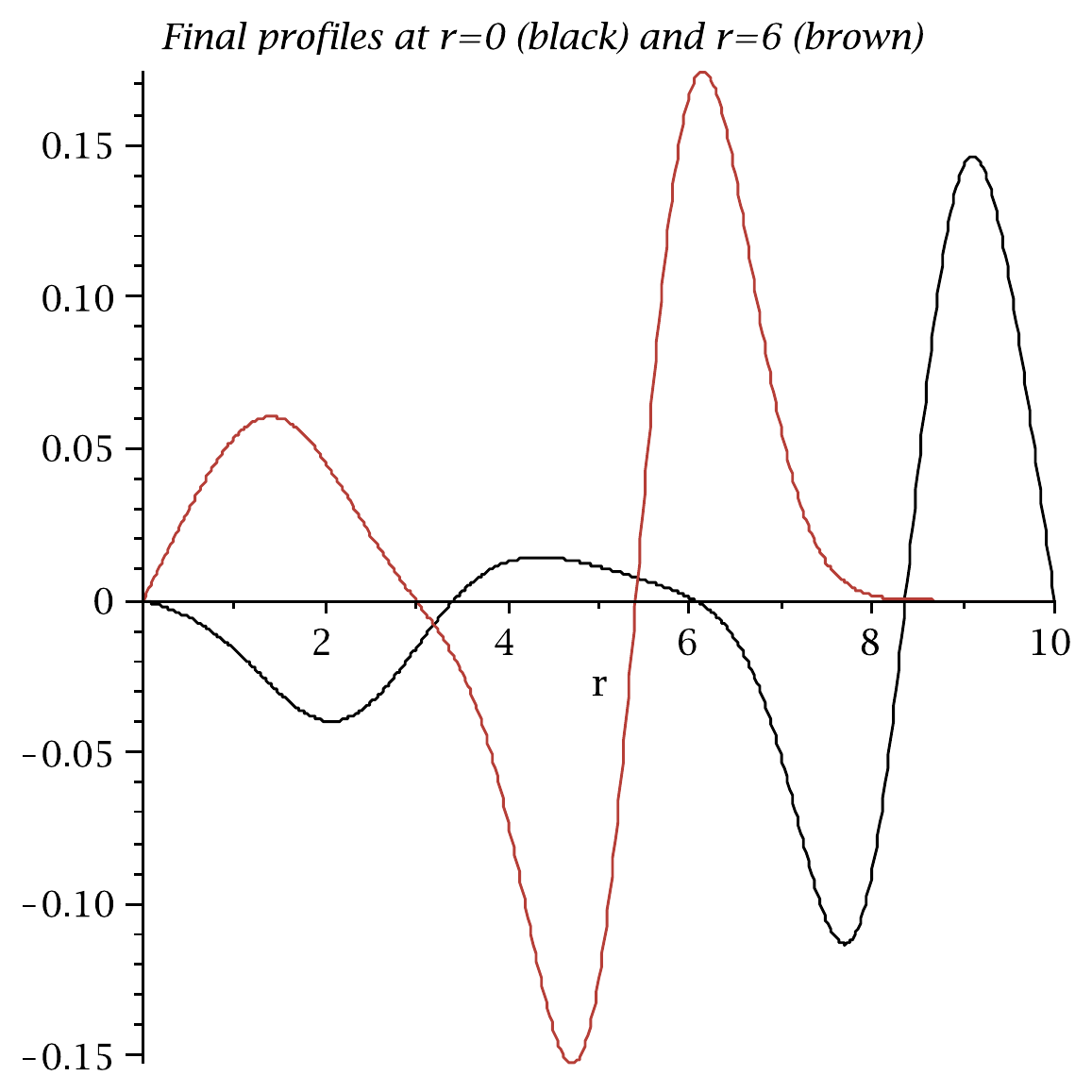}
\includegraphics*[height=5.cm,width=5.cm]{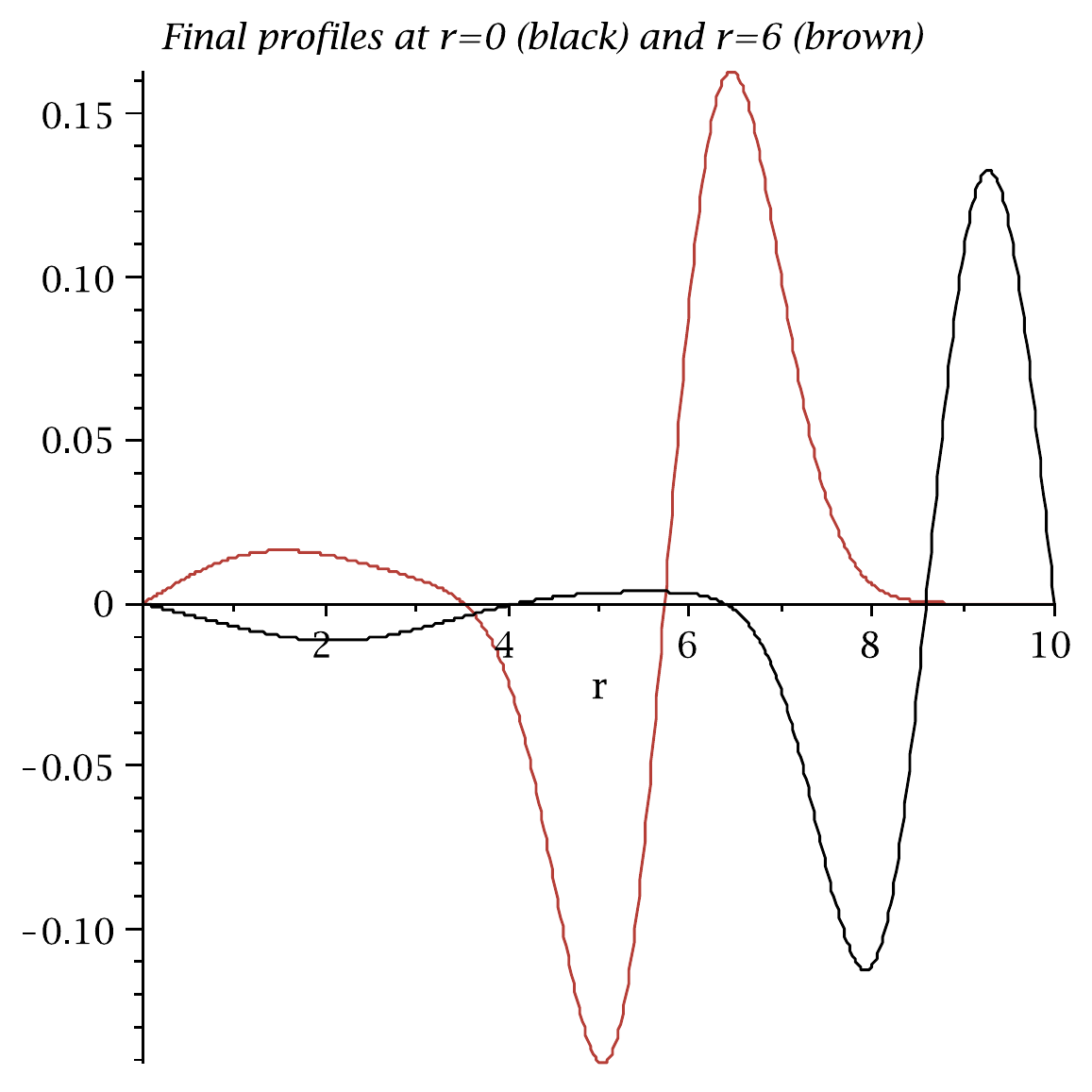}
\caption{(Color online) A comparison between the final  profiles,
in the two cases studied. The freezeout is assumed to be at
15 fm/c at the center, r=0, and 11 fm/c at r=6. 
(a,b,c) are three variants of the time dependence shown in the
precious figure. Note how the 
 amplitudes of the secondary waves gets smaller as the time variation
of speed of sound gets more smooth.}
\label{fig_final}
\end{center}
\end{figure}

We have further studied how the amplitude of the secondary wave
depends
on the variation of speed of sound. For this purpose we have
used e.g. 3 scenarios depicted in Fig. \ref{fig_cs}.
In Fig.\ref{fig_final} (a-c) we show the final profile
of
the wave at freezeout, for two histories corresponding to $r=0$
(the fireball center) and $r=6\, fm$ (fireball rim). 

Let us start with (a) in which one finds $r\epsilon$ of the secondary
wave at the rim of about the same magnitude as the primary wave.
It means that the sound intensity (which scales as $1/r^2$)
is for the secondary wave about an order of magnitude larger.
This would make the secondary wave much more likely candidate to be
observable.

 Furthermore, the azimuthal angular width of the ridge observed experimentally
is less than 1 rad, see Fig\ref{fig_peakwidth}.
 Given the magnitude of the  rapidity of the radial flow
-- up to 0.7 near the edge of the firebal
 and known thermal spread of
secondaries at freezeout, one can  explain the observed width
$assuming$  the spot size remains small as compared to the
fireball radius. This condition is
fulfilled only for the secondary wave, while for the primary one
one should include extra spread of the radial flow directions.
Thus  the secondary wave offers possible resolution of the puzzle.

 As one can see from
the solution shown in Fig. \ref{fig_final} (b,c), the effect is
unfortunately not
quite robust: the amplitude of the secondary wave decreases
if  the speed of sound is  changing more smoothly.

\section{Dual Magnetohydrodynamics (DMHD)}

   Magnetohydrodynamics is a well known part of plasma physics,
developed by Alfven, Fermi, Chandrasekhar
and many others 
 since 1940's,
see  standard textbook such as \cite{LL}. It is an approximation which keeps
only  magnetic field in Maxwell eqns, while the electric field is assumed to be totally screened. Ideal MHD approximation is 
 the limit of  $infinite$ conductivity of plasma
$\sigma\rightarrow\infty$, similar to $zero$ viscosity approximation for ideal hydrodynamics. In MHD the coupling between the field and 
and matter is obtained by inclusion of the
  (magnetic) field contribution into the stress tensor of the medium.

 The major new idea we put forward in this work is the suggestion   
 to use the {\em dual 
 Magnetohydrodynamics} (DMHD) as an approximation effectively valid
 in the near-$T_c$ (or M)
region. This proposal would of course lead to multiple
 consequences, of which we will discuss only the simplest ones.
Qualitatively, the main effect of the electric (dual-magnetic) fields in plasma
can be incorporated simply by including the fields stress tensor 
together with that of the plasma.  It would lead to extra ``elasticity" (pressure),
  helping the overall expansion a bit.  Furthermore, as fields are directed
  longitudinally, one gets certain anisotropy of the medium, with the speed
  of perturbation depending on its angle relative to the beam (and field) axis.

The notations we are going to use are simply dual to standard
ones in MHD. Thus the ``magnetic current'' of moving monopoles would be
denoted by $\tilde {\vec j}$, the coupling constant $\tilde g$,
and the (gluo)electric field $\vec E$ related to dual $\tilde {\vec H}$, but with
different normalization. The pressure of the field is
\be  p_E={\vec E^2\over 2}= {\tilde{\vec H}^2 \over 8\pi}\ee
where we keep different $4\pi$ in normalization,  Thus, apart
of tildas, Maxwell eqns look familiar, same as in textbooks.
In the infinite conductivity limit, those can be written as two 
dual-magnetic eqns
\be div\tilde{\vec H}=0\\ \nonumber
  {\partial \tilde H \over \partial t}=
 curl [\vec v \tilde{\vec H}] \ee
complemented by  Euler eqn of hydrodynamics
\be  
 {\partial \rho v_i \over \partial t}=- {\partial \Pi_{ik}
  \over \partial x_k} \ee
\be \Pi_{ik}=\rho  v_i  v_k+p \delta_{ik}-{1\over 4\pi} \left(\tilde H_i\tilde H_k-(1/2)\tilde{\vec H}^2 \delta_{ik}\right)
\ee
appended by the magnetic stress tensor, as well as the usual matter
continuity eqn 
\be{\partial \rho \over \partial t}+{\partial \rho v_i \over \partial
  x_i}=0 \ee

Thinking about possible applications of DMHD, one should first 
 clearly separate  two opposite limits: (i) the  weak field case, with $p_{field}\ll p_{plasma}$;
 and (ii) the strong field case, with $p_{field}\gg p_{plasma}$.
We will discuss them subsequently. In the former case
matter properties is only weakly affected by imbedded field, while
 strong field region expands till the pressure balance is reached, expelling
plasma from it. The situations with weak ``diffuse" fields as well as strong fields, creating
flux tubes with no plasma inside, is well known in e.g. solar plasma.

\subsection{Perturbations in the (dual)-magnetized plasma}

   In this work we restrict ourselves to the simplest problem, that of propagation of small-amplitude waves.
   in the simplest geometry, in which
only one (longitudinal or $z$) component of the field 
is nonzero. 
   Assuming the field to be permanent and homogeneous of some amplitude $\tilde{B_0}=const(x,t)$ one can  linearize
   the MHD equations,  look for plain wave solutions and determine the dispersion relation for the waves.

Qualitatively it is not hard to tell what is going to happen: depending on the relative sign of the 
pressure perturbation and that of the field, there would be $two$ solutions, one with a speed large and one smaller
than the sound speed. 
 This is well documented textbook problem, see e.g. chapter 69 of \cite{LL}, so we just mention the
 final expression for the velocities  of two ``magnetosounds" or Alfven waves
is
\be u^2_{\pm}=({1\over 2}) \left[  {\tilde{H}^2 \over 4\pi \epsilon} + c_s^2 \right]  \ee
$$ \pm   ({1\over 2})   \left[    ({\tilde{H}^2 \over 4\pi \epsilon} + c_s^2)^2 -  
 {\tilde{H}^2 cos^2\theta c_s^2   \over \pi \epsilon             } \right]^{1/2}            $$ 
where $c_s,\epsilon$ is the  speed of sound in ``unmagnetized"  medium without field and the energy density, $\theta $ is the angle between the field strength  and the
the direction of the wave propagation. 
Note a case in which  the wave goes transverse to the
field (  $cos\theta=0$ ) in which the lower mode has zero speed.

 In the case of jet quenching -- when the jet direction is more or less up to 
experimentalist to pick --  general shape of these waves can be complicated. However when the jet (or original charge fluctuation)
propagates longitudinally, in the same direction as the field, the problem is axially symmetric and results in general
in two cones. The angles of their propagation can be obtained from the
previous expression, in which  the l.h.s. is substituted by Mach relation $u\rightarrow cos(\theta)v$, v is the velocity of the jet, and solve it for the $cos\theta$.
The resulting equation  can be solved analytically, giving
\be cos\theta={\sqrt{{\tilde{H}^2 \over 4\pi \epsilon}(v^2-c_s^2)+c_s^2 v^2 } \over v^2}\ee
or zero $\cos\theta=0$ or no solution. One obvious condition is that a jet should be ``supersonic" $v^2>{\tilde{H}^2 \over 4\pi \epsilon}$.

\begin{figure}[t!]
        \includegraphics[width=8cm]{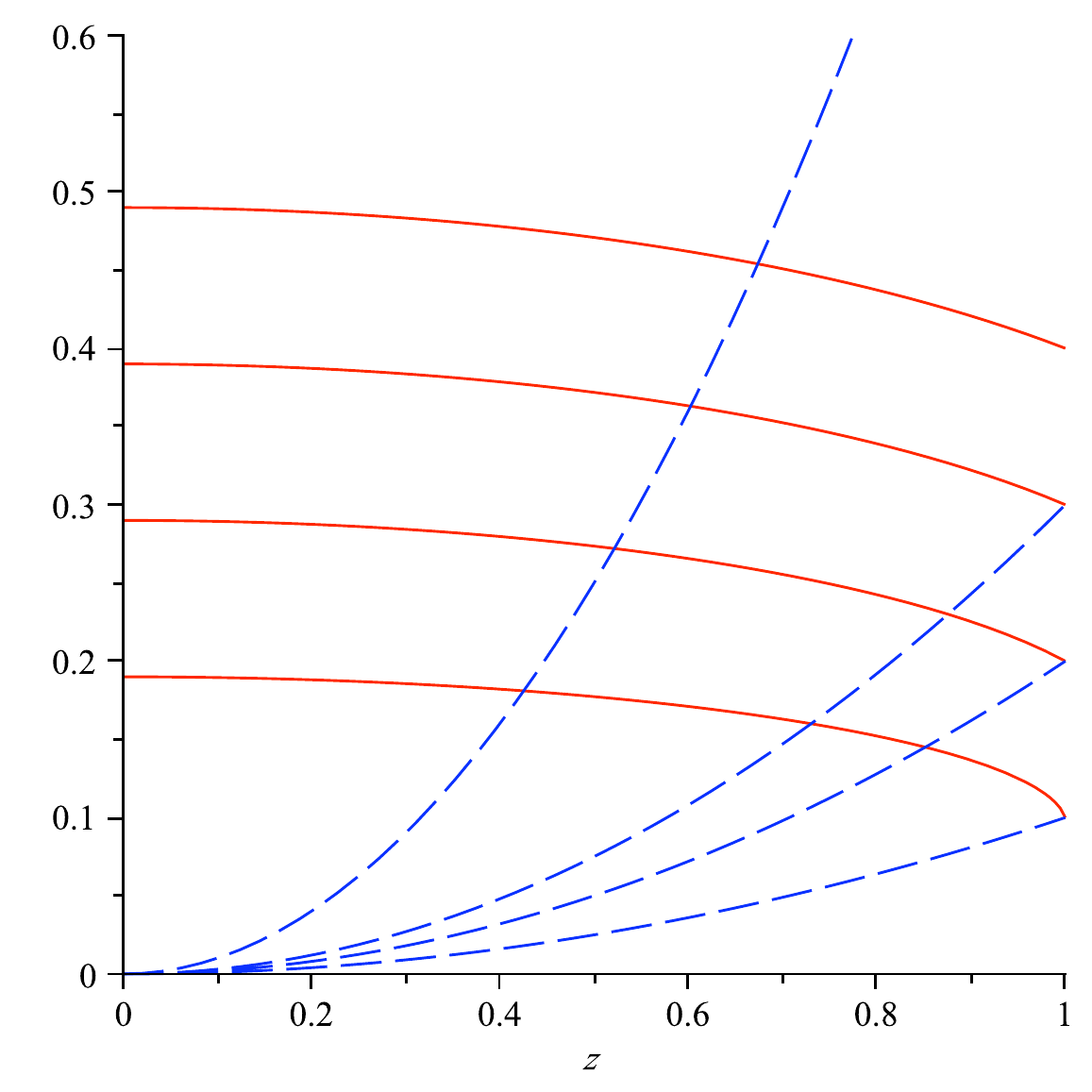}
      \includegraphics[width=8cm]{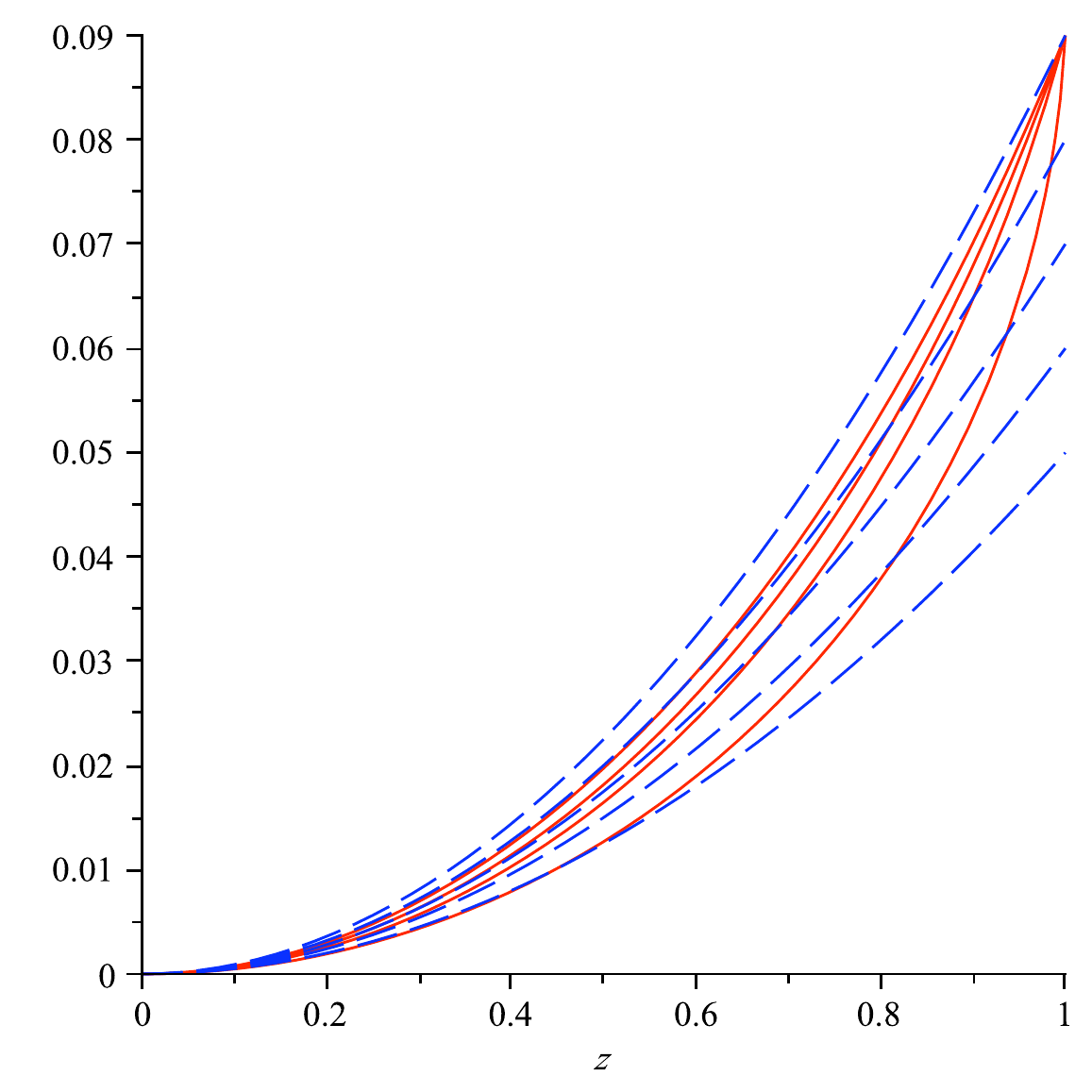}
   \vspace{-0.1in}
   \caption{
   \label{fig_twocones} (Color online)
 The solid (red) lines in figs (a) and  (b) are for the velocity of the outer  and inner cones, respectively.  The dashed (blue) lines are $v^2*z^2$ versus $z=cos(\theta)$.
 The crossing of lines  -- correspondence of the Mach condition to the wave velocity -- give the angle of the conical waves.  All curves are for $c_s^2=0.09$,
 which is the sound velocity at $T_c$ according to the lattice data. In (a) 4 solid curves are for $\tilde{H}^2 / 4\pi \epsilon=0.1, 0.2, 0.3, 0.4$, top to bottom, and 4 dashed 
for $v^2=0.1,0.2,0.3,1.$, from bottom up. In (b)  $\tilde{H}^2 / 4\pi \epsilon=0.1, 0.2, 0.3, 0.5$ for solid lines and  $v^2=0.05, 0.06, 0.07, 0.08,0.09$. }
\end{figure}

 Better insight into this equation is provided by Fig.\ref{fig_twocones} 
which display several values of the field energy density relative to matter and jet velocity $v$ for which nontrivial
cone angles exist. As one can see from the lower plot, only for rather slow jet there are  solutions with nonzero $cos\theta$: otherwise the cone
solution is at $cos\theta=0$ corresponding to zero velocity or non-expanding (stabilized) field region.
In general, note that the scales on two figures are different: the angle of the outer cone
is larger than for ordinary Mach cone, while the inner one is small if not zero.
 Therefore, qualitatively we return to the 
picture depicted in Fig.\ref{fig_sketch}(c). For obvious reason, one may think that  the slower waves  are brighter:
perhaps we observe those. 

\subsection{Flux tubes } 

In the previous subsection we considered the electric field to be constant in space. If instead we have a spot of such field, localized in transverse
 plane, inevitable there is nonzero $\partial \tilde{H}_z/\partial r$, which is a part of $curl(\tilde{H})$ and by Maxwell equation   
it should be proportional to (dual) current $\tilde{j}_\phi$. 
This tells us that a flux tube solution must have a ``coil" with a current running around and trying to cancel the field outside the spot. Ideal
DMHD has axially symmetric solution -- the classical flux tube -- similar to what is used in solar physics. 
Two nonzero equations for $r$-dependent $p(r)\tilde{j}_\phi(r) ,\tilde{H}_z(r)$ of the set are
\be \tilde{j}_\phi(r)  \tilde{H}_z(r)={d p(r) \over d r} \ee
\be  \tilde{j}_\phi(r) =-{1\over 4\pi}{d\tilde{H}_z(r)\over d r}  \ee
from which it follows that the pressure is balanced in a simple way
\be p(r) +{  \tilde{H}_z^2(r)\over 8\pi}=p(r=\infty)\ee
Two equations for three functions mean that there is functional freedom to select the tube profile.
Dissipative terms proportional to viscosity and (1/conductivity) can also be accounted for: they lead to more equations.

Furthremore,  MHD flux tubes in electromagnetic plasmas is different from that of the plasma of monopoles under considerations: unlike electrons and ions,
the monopoles and antimonopoles of opposite charges have the same mass. Thus the monopole current 
actually consists of two components of different charge, counterrotating in the opposite directions. If the magnetic plasma
is sufficiently strongly coupled, their mutual rescattering would produce significant friction and short lifetime of such configuration.   
However the main distinction between the macroscopically large flux tubes, which can be described by MHD equations, and the QCD flux tubes 
is that the latter have microscopic transverse size, smaller or comparable to the mean free path in the medium.  

  Let us at this point remind the reader brief history of flux tubes in QCD, also known as QCD strings.
 Early ideas that such flux tubes
are surrounded by 
a magnetic supercurrent due to presumed ``dual superconductivity''
in the QCD vacuum \cite{'t Hooft-Mandelstam} were refined in magnetic effective theory \cite{Baker:1996rv} and confirmed by multiple lattice studies 
such as \cite{Bali:1998de}.   With the advent of ``magnetic scenario"  \cite{Liao_ES_mono,Chernodub:2006gu}
for the near-$T_c$ region, it was suggested that 
``normal" (Bose-uncondenced)
 magnetic quasiparticles would also
%
%
%
be able to create a ``coil'' around electric flux tubes, sufficient to stabilize them.
 above certain density.

Stability condition of microscopically small  metastable flux tubes, created by
scattering of monopoles on the electric flux, has been  worked out
in two papers by Liao and myself \cite{Liao:2007mj,Liao:2008vj}.
If the flux tube is small and
monopoles can penetrate into it, their contribution to the positive or negative current depends on the (cylindrical) partial wave: as a result
expression for the current $\tilde{j}$ are rather involved.
There is no need to describe these calculations
here: let me just quote the final
 condition for the mechanical stability
 of the flux tube 
\be \label{critical_NR}
\left ( \frac{\pi M^2 c^2 k_B
T}{\hbar^2 g^2(T) n(T)} \right )^{1/2} \le 0.13 \ee
where $M(T),T,g(T), n(T) $ are the monopole mass, the temperature,
the magnetic coupling and the monopole density, respectively.
The numerical value in the r.h.s. follows from numerical solution
for the flux tubes subject to quantum scattering by magnetic 
monopoles. 

\begin{figure}[h]
\begin{center}
\resizebox*{!}{7cm}{\includegraphics{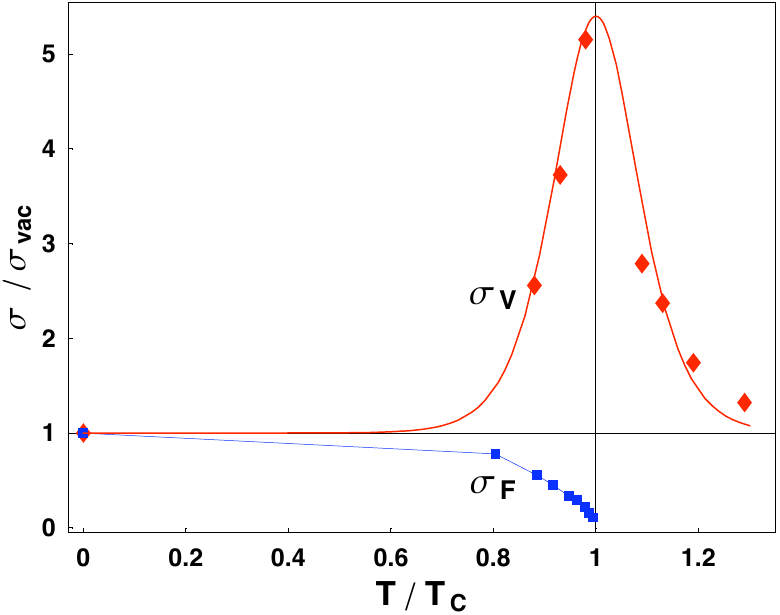}}
\end{center}
\caption{\label{fig_tension} (Color online) Effective string tensions
in the free energy $\sigma_F(T)$ (squares,from
\protect\cite{Kaczmarek_pure_gauge}) and the potential energy
$\sigma_V(T)$ (diamonds,
extracted from \protect\cite{Kaczmarek_Zantow})}
\end{figure}

So, what is the $T$ range in which the density of both condensed and ``normal" monopoles is sufficient to support the flux
tubes?  Are there any phenomenological or numerical evidences that such flux tubes actually exist?
In Refs\cite{Liao:2007mj,Liao:2008vj} the main input idea was based on lattice data of the interquark potentials at finite $T$.
In brief, the central observation is large difference between  the free energy $F(T,r)$  and the potential energy
\be V(T,r)=F(T,r)+TS(T,r) \ee
associated with quark pair at distance $r$. We will not show the potentials themselves but just their
effective string tensions for both, calculated as a slope of the linear part extracted from
 \cite{Kaczmarek_pure_gauge} , is shown in Fig.\ref{fig_tension}.
The physical difference between the two, first discussed by Zahed and myself \cite{Shuryak:2004tx} in the context of hadronic spectroscopy at finite $T$,
is that $F$ corresponds to adiabatically slow motion of the quarks, slow enough to produce maximal entropy possible and reach thermal
equilibrium at any $r$. However when quarks are moving with certain velocity $v$ away from each other, only a fraction $x$ of the
maximal entropy can be produced (because of Landau-Zener argument on level crossing): thus the effective potential would be
\be  V_{eff}(T,r)=F(T,r)+(1-x)TS(T,r) \ee 
For relatively rapid motion in which no entropy is produced, $x=0$, and one returns to $V(T,r)$.
 These arguments have been  
important to discussion of charmonium survival at RHIC as well as say dominance of baryons  \cite{our_suscept} in the near-$T_c$ region seen on the lattice.

The difference between the two potentials is discussed in detail
in  \cite{Liao:2008vj}: in short $F(T,r)$ has been related to ``supercurrent
coil'' and  $V(T,r)$ to ``normal metastable  coil''. If so,
 large peak at $T_c$ of the tension of $V(T,r)$
 is related \cite{Liao:2007mj,Liao:2008vj}
 to a peak in ``normal" monopole density at $T_c$. The condenced and normal monopole density needed to accomodate flux tubes with such
 tensions have been derived in Refs\cite{Liao:2007mj,Liao:2008vj} and compared with direct lattice observations of monopoles.
 The summary of those studies is that metastable flux tubes 
seem to exist   
at   $T<1.4T_c$, changing from stable to metastable around $T_c$.

\subsection{Production and breaking of the flux tubes}
\begin{figure}[h]
\begin{center}
\resizebox*{!}{7cm}{\includegraphics{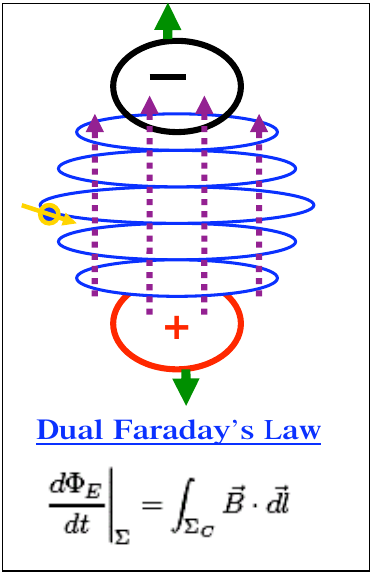}}
\end{center}
\caption{\label{fig_dual} (Color online)
Schematic demonstration of magnetic solenoidal by Dual Faraday's
law, see text.\newline}
\end{figure}

Since the cones and ridges are all created by electrically
charged particles propagating in the plasma, they do start
with rotating 
magnetic field related to ``dual Faraday'' effect shown
in Fig.\ref{fig_dual}.  While in electric plasma the acceleration happens radially,
the magnetic objects are accelerated by the magnetic field.
For a relativistic charge those are circles concentrated in a pancake of the width $\delta t\sim 1/\gamma\sim \sqrt{1-v^2}$ with the spatial distribution
\be  \vec B \sim {\gamma g_e [\vec v \vec r_\perp] \over [\gamma^2(z-v t)^2+r_\perp^2]^{3/2} }\ee
Monopoles (with magnetic charge $g_m$) experience instant kick, in the corresponding direction. Its magnitude
has no gamma factor and is proportional to the product of electric and magnetic couplings
$ g_eg_m /4\pi=n$, a Dirac integer. 
(For elementary quark/gluon
electric charge and   for Polyakov-t'Hooft
monopoles it is just ).  Of course electric charge fluctuations
can lead to larger charge values as well.  This velocity  (near-)instantaneously produces a  current
(or  a ``coil')  running around the flux.  The questions under which condition
it is robust enough to contain the field into a static flux tube  will be
discussed elsewhere.

  The interpretation of the linear part of the potential energy coming from the lattice $V(T,r)$ 
as a metastable flux tube mentioned in the previous subsection,   Fig.\ref{fig_tension},
%
can be used to provide estimates of the absolute 
amount of energy/matter involved. At $T_c$ the values are astonishingly
 large: the energy  per unit length (tension) reaches at its peak
$5\, GeV/fm$, corresponding to the entropy density
of about $30 fm^{-1}$. One may ask if those parameters can be compatible with
observations of the ridges/cone at RHIC.

Let me start addressing this issue from the point of view of energy first. The total energy of the flux tube, or
the work done by its tension on the departing charges (large-$x$ valence quarks) is $\delta E=\tau\sigma$.
For the M-phase lasting $\tau\sim 5 \fm/c$ and the proposed V-potential tension, one gets the energy loss of about 25 GeV.
Keeping in mind that the total energy of each nucleon in the center of mass is 100 GeV, at  RHIC energy used for heavy ions, 
and that each nucleon has three valence quarks plus sea plus gluons, we conclude that it is comparable to quark total energy.
So, if the string would not break, it would be able to transfer valence quarks from the fragmentation region to  midrapidity: and we know
from experiment that it happens with very small probablity.

The conclusion is then that such strings $must$ get broken at time shorter than $\tau\sim 5 fm/c$ used in this estimate. This is
also known to be true for the usual (vacuum or $T=0$) strings: for those the rate of breaking has been phenomenologically
extracted from hadron decays, especially of hadrons with large angular momenta (Regge trajectories) and
event generators based on the Lund model.   

Naively, one might think that ``metastable" flux tubes in the M-phase should have  $smaller$ lifetime than the vacuum QCD strings.
First, their decay can be related not only to (i) the quark pair
production, as in vacuum, but also
to (ii) just picking up quarks from the ambient matter.
However, looking at numbers more closely one finds that it is not nacessarely so.

Quark pair
production is described by Schwinger
fermion pair production rate  (the leading exponent only) in a constant electric field $E$
\be {dW \over d^4x}={e^2 E^2 \over \pi^2} 
exp\left(-{\pi M_q^2 \over e E    }\right) \ee
where  $M$ is the charged particle mass.
The tension scales as $\sigma \sim E^2  r_\perp^2$, where $E,r_\perp$ are 
electric field and transverse size of the flux tube. There is a universal flux, so $E r_\perp^2=const$,
and thus in any change of the tube the field scales as the first power of the tension. If the
tension grows by some factor, e.g. $\sigma(T_c)/\sigma(0)\sim 5$ as suggested by lattice potential $V$,
the field should grow by the same factor. Naively, it would greatly reduce its lifetime\footnote{In 1990's people discussed the so called ``color ropes",
as flux tubes with larger flux and thus the field strength, as compared with the vacuum strings. It was done to
explain higher strangeness content in AA relative to pp collisions: but the lifetime of such ropes should be smaller than that of the elementary strings. }.

Yet the effective mass of quark quasiparticles near $T_c$ is also quite different from ``constituent quark mass'' in the vaccum  ($T=0$)
 $M_q\approx
330 \, MeV$. 
 For example, direct lattice
studies \cite{Petreczky:2001yp} show that $M_q,M_g\sim 800\, MeV$ at $T=1.5Tc$, with perhaps even larger value at $T_c$.
Thus the combination which enters the exponent of the Schwinger formula $M^2_q/\sigma$ is not decreasing but rather grows.  
It suggests a rather counterintuitive conclusion: flux tubes at $T_c$ may have $smaller$ breaking rate than in the vacuum.

The same argument, based on large quark/gluon mass near $T_c$, leads to small Boltzmann weight  $exp(-M/T)\ll 1$
 which explains small density of these quasiparticles and thus small probablity of a 
string breaking by picking up free charges.

\begin{figure}[h]
\includegraphics[width=6.cm]{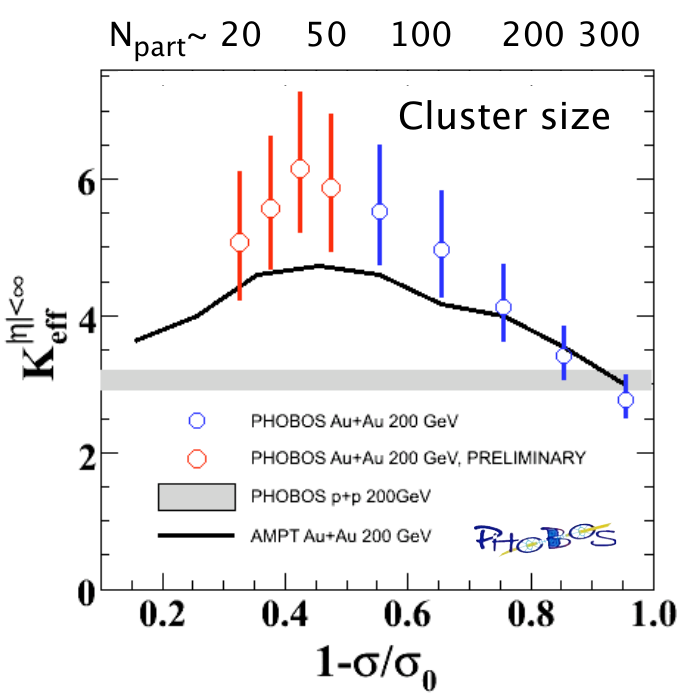}
\includegraphics[width=6.cm]{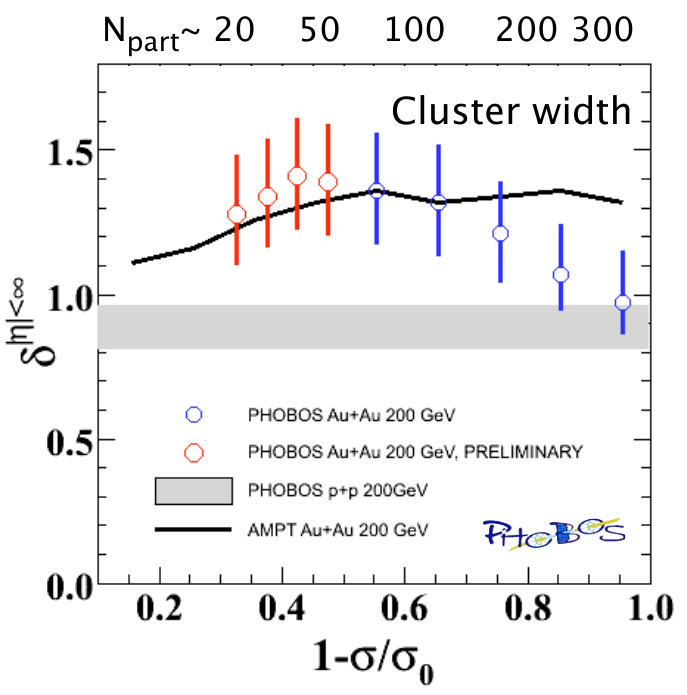}
\caption{(Color online) The cluster size (a) and width (b) as a function of centrality (cross section fraction) in AuAu collisions
at RHIC full energy, from PHOBOS \cite{Stephans}. The size is the number of charged particles associated with the cluster and the width
is in rapidity. 
}
\label{fig_clusters}
\end{figure}

Remarkably,  recent RHIC data provided direct experimental indications for enhanced stabilty of the flux tube in matter relative to pp. 
We will use those  from PHOBOS collaboration, which has large rapidity
coverage of their silicon detector.  Fig.\ref{fig_clusters}(a) shows that the number of charged
particles in a cluster observed in AuAu collisions (points) is about twice that seen in correlation studies of the pp collisions (shaded horizontal region),
so that they reach the size of 6 charged particles (or 9 total).  
(The exception are central collisions on the right, in which case extensive hadronic after-burning kills the correlations.) 
The figure (b) shows that the produced clusters are not near-isotropically decaying resonances as in pp, (shaded horizontal region), but
are instead more extended in rapidity. This last fact shows their
direct relation to  the ``soft ridge" and flux tubes. Similar CuCu data (not shown) demostrate basically the same clusters, provided
the same centrality is taken: this shows that geometry and surface-to-volume ratio is important. 
 Taken together, we interpret those clustering data as direct proof of significant changes in the flux tube decay parameters in AuAu relative to pp: 
the tubes apparently gets denser and decay less frequently, into larger pieces.

Let me also try to connect the multiplicity in the clusters observed with the entropy as seen in lattice potentials.
At the peak the entropy per length is as high as $dS/dL=5 GeV/fm/T_c\sim 30$.   
If this entropy all goes into final pions at freezeout, one can estimate the cluster absolute length.
Neglecting the pion mass -- that
is considering matter to be ideal massless bose gas at freezeout -- one find that entropy is 3.6 per particle, or $S\sim 36$ for 10 pions in a cluster.
 Thus if the decay takes place in the M phase, the cluster corresponds to a
string length of 1.2 fm or so. However if the decay rate is small and the string survivies through the M-phase to hadronic phase,
the string tension is reduced to the vacuum value 1 GeV/fm, the  length corresponding to the observed cluster would be as long as 6 fm.
In principle this information
 can be directly related
to the r.m.s. rapidity width
 of the cluster  seen from fig.\ref{fig_clusters}(b).   We are planning to compare it with some PYTHIA-like simulations to tell whether
 those clusters can be consistently reproduced.
 
 \subsection{Discussion and predictions }
 In this section we have suggested two consequences of the small electric screening mass in the M-phase (near $T_c$ region):
 (i) unscreened bulk electric fields and (ii) metastable (and possibly even relatively long lived) electric flux tube. Taken together,
 they were called ``QGP corona", with reference to similar phenomena in solar plasma. We had tried to connect this picture with
 phenomenology of fireball perturbations, namely ridges and cone.
 Now we would like to speculate further along this line, pointing out some further consequences of the ``corona" idea, which can
 be further tested in experiment. 
 
 One direction is related with the 
 predictions of the energy/centrality dependence of the ridges and cones. Already in the introduction we have shown in our sketch of the idea,
 depicted in Fig.\ref{fig_dual_mag}, we indicated that since the geometry of the M-phase domain is quite different for different collision energies. For RHIC
 the M-phase sits on the outside of the fireball (Fig.(a)), thus it get a maximal boost from the  hydrodynamical expansion and may produce
 ridges with rather narrow peaks in the azimuthal angle. For much lower energies -- corresponding both to the SPS fixed target experiments and
 planned RHIC scan down) --   one should find the M-phase only on the inner part of the fireball (Fig.(b)), which experience little hydro boost if at all.
 The logical prediction is then that no soft or hard ridges should be observed in this case. Standard modelling using the realistic geometry/density
 distributions should make those predictions quantitative.
 
 The ``cone" is a different story: first of all, those do not rely on overall hydrodynamics and thus may well happen at the very 
 center of the firball and yet be observed. Second important distinction:
 ``cones" are perturbations created by the ``away-side" jet, and thus have completely different geometry/timimg. If the trigger jet is surface biased,
 the away side jet has to fly through rather long path inside fireball, with its length varying roughly between its radius and diameter, 6-12 fm.
 As one can see from hydro solution, 
 the M-phase starts at time zero at the edge and time about 5 fm/c at the center.
 Combining two observations together, one would see, that the away-side jet is travelling most of its long path in the M-phase. Furthermore,
 even at low collision energies, when the M-phase occupies only the central part of the fireball, this conclusion is still fulfilled. The prediction
 then is that ``cones" should not show very strong dependence on collision energy and centrality, in contrast to ridges.  

Let us now see how these ideas confront the available data. SPS experiments -- CERES and NA49 -- have not seen anything like ridges.
The centrality and energy dependence of RHIC data we shave shown in Fig.\ref{fig_transition} do indeed suggest that this phenomenon is disappearing
rather rapidly. On the other hand,  both CERES and NA49 observe away-side structures which are remarcably similar to RHIC
data on the away-side, see review by Harald AppelshŠuser at the workshop \cite{cathie_workshop}.  We conclude that the picture in which ridges
originate from spacial part of the ``QGP corona", while cones are from its temporal part, at times 5-10 fm, is in qualitative agreement with the data. 

Finally, let me indicate one more direction of future studies: possible role of the unscreened electric fields in the M-phase in early-time hydro evolution.
The pressure of the field, adding to (very low) pressure of the M-phase may help to start hydro a bit earlier and help explain the HBT puzzle.  Some studies of the kind 
(but with non-equilibrium fields rather than DMHD ones) have been made in Ref.
 \cite{Vredevoogd:2008id}.

\section{Conclusions }

In summary, we discussed two scenarios of the evolution of extra
energy/matter deposited by some fluctuations
on top of the ``Little bang''. 
In the scenario (A) we solved equations for
propagation of sound 
 with variable speed of sound  in expanding matter, using
the Hubble flow
approximation.
We have found
that the rapid drop of the sound velocity in the ``mixed''
phase generates the secondary wave, which under certain conditions may 
 be brighter and smaller
in size, and thus is much better candidate for
the observed ``cone'' and ``ridges''.
 However this effect strongly depends on how sharply
 the speed of sound changes near $T_c$,
 with the secondary wave being washed away  if 
changes in the speed of sound gets are too  smooth.

In the scenario (B) 
we  assumed that electric field remains unscreened for the duration of the M-phase (near $T_c$) and used dual Magnetohydrodynamics.
We   also found a potential for two expanding cylinders/cones. Furthermore,
the speed of one mode can be zero which means that the flux tube
 of electric gauge field can be pressure-
stabilized, by a (metastable) ``coil'' or current by magnetic charges
in plasma. We have argued that this phenomenon is  likely
to stabilize the flux tubes
 in the near-Tc region, approximately at $.8Tc<T<1.4Tc$.
However this temperature interval does not account to all the time
of the ``Little Bang'' at RHIC energies. Presumably the ``coil''
effect is still present at higher $T$, 
partially reducing the tube expansion.

For the readers who may be surprised by similarity of the ``double cones", let us remind that similar phenomena 
are well known in other fields of physics and may appear for multiple reason. In particular, in ``dusty" strongly coupled electrodynamic plasmas
double Mach cones have been experimentally observed (see e.g.\cite{dusty}): in this case these are sound and ``shear" or hydroelastic modes which generate them.
 (Perhaps this option can lead to a ``scenario C" not yet considered.)

The very fact that we discuss two competing scenario should tell the reader that 
at the moment it is hard to tell whether they are robust enough to survive further scrutiny and
 explain pertinent observations. 
 The ``acoustical solution'' is  not quite robust,
 it does work only for rather sharp QCD transition which current lattice data do not support.    
  Survival of electric field in the QGP corona is to be studied more, as well as 
presence of metastable flux tubes.
Clearly more  theoretical
work is needed, including dedicated lattice studies
of both monopoles and speed of sound.
As far as experiment is concerned, it would be highly
important to insure planning so that
 that jet correlations in question 
can be followed with sufficient accuracy during the
planned RHIC scan toward the lower collision energies.
Changing the collision energy one  changes the timing of the fireball eras, 
 in a predictable manner, thus for the phenomena we proposed 
 some quantitative predictions can be worked out and tested. 

Having said that, we emphasize that those theoretical and experimental studies
seem to be very much justified as both  may potentially  become  
important discoveries.
 If the ``acoustic scenario'' (A) is the explanation, it would be
a direct experimental 
signature of  sharp quasi-1st-order
 QCD phase transition.
If the DHMD scenario (B) would be confirmed, it
would  be quite significant finding,  
confirming  reality  of ``QGP corona" in which physics is different from what it is inside, as much as it is on the Sun.
It will be a big boost to ``magnetic scenario" for the
 near-$T_c$ region     
\cite{Liao_ES_mono,Chernodub:2006gu}.

\vskip .25cm {\bf Acknowledgments.} \vskip .2cm
The author indebted to his collaborators (and former students), JinFeng Liao and Jorge Casalderrey-Solana,
with whom the development of some of these ideas have been made.
I am also indepteed to Larry McLerran and Raju Venugopalan,
who attracted his attention to the soft ridge problem, to
 participants of the correlation workshop \cite{cathie_workshop}
for illuminating discussions, and also to G.Stephens for his explanation of PHOBOS correlation data. 
This work was supported in parts by the US-DOE grant
DE-FG-88ER40388.

\end{document}